\documentclass{jfm}
\usepackage{graphicx}
\usepackage{epstopdf, epsfig}
\usepackage{bm}
\usepackage{amsmath}
\usepackage{amssymb}
\usepackage{soul}
\usepackage{subfigure}
\usepackage{caption}
\usepackage[dvipsnames]{xcolor}
\usepackage{verbatim}

\newcommand{\blue}[1]{\textcolor{blue}{#1}}

\newcommand{\nc}[1]{{\color{RedViolet}{#1}}}

\shorttitle{Lattice Boltzmann model for multicomponent mixtures}
\shortauthor{N. Sawant, B. Dorschner and I. V. Karlin}

\title{Consistent lattice Boltzmann model for multicomponent mixtures 
}

\author{N. Sawant, B. Dorschner \and I. V. Karlin \corresp{\email{ikarlin@ethz.ch}}}

\affiliation{Department of Mechanical and Process Engineering, ETH Zurich, 8092 Zurich, Switzerland}

\begin{document}
	
	\maketitle
	
	\begin{abstract}

A new lattice Boltzmann model for multicomponent ideal gas mixtures is presented.  The model development consists of two parts. First, a new kinetic model for Stefan--Maxwell diffusion amongst the species is proposed and realized as a lattice Boltzmann equation on the standard discrete velocity set. Second, a compressible lattice Boltzmann model for the momentum and energy of the mixture is established. 
Both parts are consistently coupled through mixture composition, momentum, pressure, energy and enthalpy whereby a passive scalar advection-diffusion coupling is obviated, unlike in previous approaches.
The proposed model is realized on the standard three-dimensional lattices and is validated with a set of benchmarks highlighting various physical aspects of compressible mixtures. 
Stefan--Maxwell diffusion is tested against experiment and theory of uphill diffusion of argon and methane in a ternary mixture with hydrogen. The speed of sound is measured in various binary and ternary compositions. We further validate the Stefan--Maxwell diffusion coupling with hydrodynamics by simulating diffusion in opposed jets and the three-dimensional Kelvin--Helmholtz instability of shear layers in a two-component mixture. Apart from the multicomponent compressible mixture, the proposed lattice Boltzmann model also provides  
an extension of the lattice Boltzmann equation to the compressible flow regime on the standard three-dimensional lattice.
	\end{abstract}
		\vspace{-0.25cm}
	
\begin{small}
\tableofcontents
\end{small}
\section{Introduction}
\label{sec:intro}
 The lattice Boltzmann method (LBM) is a recast of fluid dynamics into a fully discrete kinetic system of designer particles with the discrete velocities $\bm{c}_i$, $i=0,\dots,Q-1$, fitting into a regular space-filling lattice, with the kinetic equation for the populations $f_i(\bm{x},t)$ following a simple algorithm of ``stream along links $\bm{c}_i$ and collide at the nodes $\bm{x}$ in discrete time $t$". 
 Since its inception  \citep{Higuera_1989_SucciBenzi,Higuera_1989}, LBM has evolved into a versatile tool for the simulation of complex flows including  transitional flows \citep{Dorschner2017JFM}, flows in complex moving geometries \citep{dorschner2016entropic}, compressible flows \citep{Frapolli2016prl,PonD2018}, multiphase flows \citep{Mazloomi2015prl,Mazloomi2017JFM,Woehrwag2018prl} and rarefied gas \citep{shan2006kinetic}, to mention a few recent instances; see \citet{succi,krueger} for a discussion of LBM and its application areas.
 
In view of extensive development, it seems surprising that the multicomponent gas mixtures so far resisted a significant advancement in the LBM context. Importance of compressible mixtures is hard to overestimate because they are a prerequisite for combustion applications \citep{williams}. However, incorporating even the basic mechanism of multicomponent diffusion in gases, the Stefan--Maxwell diffusion, remains an essentially unsolved problem in the LBM context, in spite of that the Stefan--Maxwell diffusion is itself a derivative of Boltzmann's kinetic theory \citep{chapmanCowling}.
It is worth reminding that the Stefan--Maxwell diffusion mechanism is well recognized as a fundamental feature of gas mixtures, supported by experiment \citep{toor1957,duncan1962,arnold1967} and molecular dynamics simulations  \citep{wheeler,krishnaAndBaten}. 
As highlighted by \citet{krishna}, the Stefan--Maxwell diffusion is more subtle than the conventional Fick's model. The latter implies that any component in a mixture moves from higher to lower concentration regions.
The Stefan-Maxwell model, on the other hand, accounts for binary interaction between each of the species pairs through pairwise diffusion coefficients and can lead to counter-intuitive effects such as uphill diffusion when a component in a ternary mixture moves from lower to the higher concentration region \citep{toor1957,duncan1962,arnold1967}. Among applications of Stefan--Maxwell diffusion in the conventional computational fluid dynamics, we mention recent studies of diffusion in fuel cells \citep{hsing,stockie,suwanwarangkul}.

However, the majority of existing LBM models for the multicomponent mixtures \citep{salvoMulti,Chiavazzo_2009,feng2019hybrid,huang2019lattice,hosseini2018,lin2017multi} are bound to use the Fick diffusion model rather than the Stefan--Maxwell. 
To the best of our knowledge, the only LBM realization of the Stefan--Maxwell diffusion was reported in a very recent work of \citet{chai}; However, the two-dimensional LBM of \citet{chai} is restricted by the isothermal and isobaric assumptions and can thus not provide a basis for the development of a compressible mixture LBM.
%
Another obstacle arises at the coupling of diffusion to the transport of momentum and energy.
The simplest way of tackling multicomponent mixtures with the LBM is by representing the dynamics of the species by a advection-diffusion equation (see, e.\ g. \citet{Chiavazzo_2009} and references therein). In this approach, the species are treated as passive scalars, advected with the fluid velocity and the species do not influence the fluid or other species. 
The passive scalar viewpoint on LBM for mixtures was adopted and extended in a number of recent proposals   \citep{feng2019hybrid,huang2019lattice,hosseini2018,lin2017multi}. However, apart from the inability of incorporating the Stefan--Maxwell diffusion, the passive scalar approach suffers from a more fundamental shortcoming, namely
thermodynamic inconsistency.
For example,  the aforementioned models do not readily recover the correct heat flux in the multicomponent system \citep{chapmanCowling,williams,bird2006transport} and miss the enthalpy flux due to diffusion.


In this paper, we revisit the LBM construction for a compressible multicomponent mixture, focusing on a thermodynamically consistent coupling between the Stefan--Maxwell diffusion and momentum and energy transfer in the system. 
We begin in Sec.\ \ref{sec:stefanMaxwell} with setting up a kinetic system for the species in the $M$-component mixture. The construction follows the path of so-called quasi-equilibrium relaxation models \citep{Gorban1994,QELBM}; see \citet{salvoMulti} in the context of isothermal mixtures. Here, we significantly extend the quasi-equilibrium kinetic model for the species to a generic ideal gas equation of state and, unlike in the earlier approach of \citet{salvoMulti}, enabling the Stefan--Maxwell constitutive relation. After a short summary of nomenclature in Sec.\ \ref{sec:thermoIdealGasMixture}, the species kinetic equations are introduced in Sec.\ \ref{sec:formulation}, in the continuous time-space setting. We show in Sec.\ \ref{sec:limitSPEC} that the proposed kinetic equations recover the Stefan--Maxwell diffusion together with the barodiffusion in the hydrodynamic limit. The species kinetic equations are realized on the standard set of discrete velocities in Sec.\ \ref{sec:27spec}. In Sec.\ \ref{sec:numericalImplementation}, we derive the lattice Boltzmann scheme for the species kinetic equations following the technique of integration along characteristics introduced by \citet{he1998}. This concludes the first part of the model development.

We continue in Sec.\ \ref{sec:lbMixtures} with a mean-field lattice Boltzmann formulation of the mixture momentum and energy. After a summary on the mixture energy and enthalpy in Sec.\ \ref{sec:thermo2}, we present a generic two-population lattice Boltzmann equation for the mixture. We note that the mean-field approach requires only two lattice Boltzmann equations, one for the  mixture density and momentum and another one for the energy. While the two-population LBM is established approach for a single-component compressible fluid \citep{frapolli2015,hosseinCompressible}, the application of the two-population techniques to the mixture requires a modification of the nonequilibrium fluxes discussed in Sec.\ \ref{sec:kinetic}. The mixture density, momentum and energy equations are presented in Sec.\ \ref{sec:hydromixture} while details of their derivation with the Chapman--Enskog analysis \citep{chapmanCowling} are summarized in the Appendix \ref{sec:ceNavierStokes}. The two-population mixture LBM is realized on the standard lattice in Sec.\ \ref{sec:realization} where we extend the two-dimensional compressible LBM of \citet{hosseinCompressible} to three-dimensional mixtures. 
Finally, in Sec.\ \ref{sec:coupling} we discuss the coupling between the $M$ LBM equations for the species  and the double-population mean field mixture LBM. The resulting LBM provides a reduced description of the $M$-component mixture with $M+2$ tightly coupled lattice Boltzmann equations, unlike a standard kinetic approach which would require $2\times M$ kinetic equations. 

In Sec.\ \ref{sec:results}, the LBM model is validated with a number of select benchmarks. After a summary of general aspects of numerical implementation in Sec.\ \ref{sec:overviewnumerics}, we present a simulation of diffusion of argon and methane ternary mixture with hydrogen in the Loschmidt tube apparatus in Sec.\ \ref{sec:ternary}, along with the classical experiment of \citet{arnold1967} and theoretical discussion of \citet{krishna}. We show that the LBM simulations reproduce in a quantitative fashion the experimentally observed features of the Stefan--Maxwell diffusion such as uphill and osmotic diffusion and the diffusion barrier \citep{toor1957,duncan1962,arnold1967,krishna}.
The coupling between  hydrodynamics and diffusion is validated in a counterflow diffusion in opposed jets in Sec.\ \ref{sec:OpposedJets} and the speed of sound measurements are presented in Sec.\ \ref{sec:soundSpeedTest} for probing the compressible flow aspect of the model.  Finally, a simulation of the three-dimensional Kelvin--Helmholtz instablity in a binary mixture is reported in Sec.\ \ref{sec:khi} as a test for the performance of the proposed LBM in a complex flow.
Conclusions are drawn in Sec.\ \ref{sec:conclusion}.

\section{Lattice Boltzmann model of Stefan--Maxwell diffusion}
\label{sec:stefanMaxwell}

\subsection{Composition and equation of state of ideal gas mixture}
\label{sec:thermoIdealGasMixture}

We begin with introducing some nomenclature and notation. Let us consider a mixture composed of $M$ ideal gases. 
The composition is described by the species densities $\rho_a$, $a=1,\dots, M$, while  the mixture density is
 \begin{equation}
 \rho=\sum_{a=1}^{M}\rho_a.
 \end{equation}
Equivalently, the mixture composition is defined by the mixture density $\rho$ and the $M-1$ independent mass fractions $Y_a$,
\begin{equation}
 Y_a=\frac{\rho_a}{\rho},\ \sum_{a=1}^{M} Y_a = 1.
 \label{eqn:mixtureDensity}
\end{equation}
With the molar mass of the component  $m_a$, the mean molar mass $m$ depends on the composition,
\begin{equation}
\frac{1}{{m}}=\sum_{a=1}^M\frac{Y_a}{m_a}.
\label{eqn:meanMolecularWeight}
\end{equation}
The equation of state (EoS) provides a relation between the pressure $P$, the temperature $T$ and the composition,
\begin{equation}
P=\rho R T.
\label{eqn:eosIdealGas}
\end{equation}
Here, $R$ is the specific gas constant that contains the information about the composition of the gas by way of the mean molar mass $m$, 
\begin{equation}
R=\frac{R_U}{m},
\label{eqn:defineSpecificGasContant}
\end{equation}
where $R_U\approx 8.314\; kJ/K\cdot kmol$ is the universal gas constant.
Thus, for a mixture of ideal gases, the specific gas constant $R$ is a function of local composition and changes in space and time.
The pressure of an individual component $P_a$ is related to the pressure of the mixture $P$ through Dalton's law of partial pressures as follows:
\begin{equation}
P_a=X_a P,
\label{eqn:dalton}
\end{equation}
where the mole fraction of a component $X_a$ is related to its mass fraction $Y_a$ as 
\begin{equation}
X_a=\left(\frac{m}{m_a}\right)Y_a, \ \sum_{a=1}^{M} X_a = 1.
\label{eqn:moleFractionDefinition}
\end{equation}

A consequence of Dalton's law of partial pressure is that $\sum_{a=1}^{M}P_a=P$. Combined with the equation of state (\ref{eqn:eosIdealGas}), the partial pressure $P_a$ takes the form,
\begin{equation}
P_a=\rho_a R_a T,
\label{eqn:eosIdeaGasComponent}
\end{equation}
where $R_a$ is the specific gas constant of the component,
\begin{equation}
    R_a=\frac{R_U}{m_a}.
    \label{eq:Ra}
\end{equation}
With these thermodynamic relations in mind, we proceed to setting up the kinetic equations that recover the Stefan--Maxwell diffusion in the macroscopic limit.

\subsection{Kinetic equation for the species}
\label{sec:formulation}

In this section, we set up kinetic equations which recover the Stefan--Maxwell diffusion model for a $M$-component ideal gas mixture.
Each component is described by a set of populations $f_{ai}$, $a=1,\dots,M$, corresponding to the discrete velocities $\bm{c}_i$, $i=0,\dots, Q-1$. 
The proposed kinetic equation for each of the  species $a$ is written as,
\begin{equation}
\partial_t f_{ai} + \bm{c}_{i}\cdot \nabla f_{ai} = \sum_{b=1}^M \frac{1}{\theta_{ab}} \left[ \left( \frac{f_{ai}^{\rm eq}-f_{ai}}{\rho_a} \right) - \left( \frac{f_{bi}^{\rm eq}-f^*_{bi}}{\rho_b} \right) \right].
\label{eqn:stefanMaxwell} 
\end{equation}
Here, $\rho_a$ is the density of the component $a$, which is defined as the zeroth moment of populations $f_{ai}$,
\begin{equation}
\rho_a=\sum_{i=0}^{Q-1}f_{ai}.
\label{eq:density1}
\end{equation}
Furthermore, a symmetric set of relaxation parameters $\theta_{ab}=\theta_{ba}$ shall be related to the binary diffusion coefficients below.
The  equilibrium $f_{ai}^{\rm eq}$ and the quasi-equilibrium $f^*_{ai}$ populations will be fully defined in 
section \ref{sec:27spec}. Here, we only need to specify the conditions for the low-order moments thereof. 
To that end, let us introduce the partial momenta $\rho_a\bm{u}_a$ as first moments of the species' populations,
\begin{equation}
\rho_a\bm{u}_a=\sum_{i=0}^{Q-1} f_{ai}\bm{c}_i.
\label{eq:ua2}
\end{equation}
The quasi-equilibrium populations $f_{ai}^*$ must satisfy the following set of constraints,  
\begin{align}
\sum_{i=0}^{Q-1} f_{ai}^{*} &=  \rho_a,
\label{eq:rhoqe}\\
\sum_{i=0}^{Q-1} f_{ai}^{*}\bm{c}_i &=  \rho_a \bm{u}_a.
\label{eq:ua}
\end{align}
The momenta of the components sum up to the mixture momentum,
\begin{equation}
\sum_{a=1}^{M} \rho_a \bm{u}_a = \rho \bm{u}.
\label{eqn:mixtureMomentum}
\end{equation}
The equilibrium populations $f_{ai}^{\rm eq}$ have to verify the following set of constraints:
\begin{align}
\sum_{i=0}^{Q-1} f_{ai}^{\rm eq} &=  \rho_a,
\label{eqn:f0ZerothMoment}\\
\sum_{i=0}^{Q-1} f_{ai}^{\rm eq}\bm{c}_i &=  \rho_a \bm{u},
\label{eq:u}
\\
\sum_{i=0}^{Q-1} f_{ai}^{\rm eq}\bm{c}_i\otimes \bm{c}_i&=  P_a\bm{I} +\rho_a\bm{u}\otimes \bm{u}.
\label{eq:Pa}
\end{align}
In Eq.\ (\ref{eq:Pa}), the partial pressure $P_a$ (\ref{eqn:eosIdeaGasComponent}) depends on the temperature $T$, which is obtained from the mixture kinetic equations of Sec.\ \ref{sec:lbMixtures}.
Finally, the quasi-equilibrium distribution must match the equilibrium if the species velocity equals the velocity of the mixture, $\bm{u}_a=\bm{u}$: 
\begin{equation}\label{eq:match}
f^*_{ai}(\bm{u}_a)\big|_{\bm{u}_a=\bm{u}}=f^{\rm eq}_{ai}(\bm{u}).
\end{equation}
Some comments are in order:
\begin{enumerate}
\item The $M$-component kinetic system satisfies  $M+D$ conservation laws, where $D$ is the space dimension: The densities $\rho_1, \dots, \rho_M$ and the vector of fluid momentum $\rho\bm{u}$ are locally conserved fields.
\item Thanks to the matching condition (\ref{eq:match}), the relaxation term on the right hand side of Eq.\ (\ref{eqn:stefanMaxwell}) vanishes only at the equilibrium.
\end{enumerate}
We now proceed with the identification of the relaxation parameters $\theta_{ab}$ in terms of the Stefan--Maxwell binary diffusion coefficients. 

\subsection{Hydrodynamic limit of kinetic equations for the species }
\label{sec:limitSPEC}

Evaluation of the zeroth and of the first moments of the kinetic equation (\ref{eqn:stefanMaxwell}) results in the balance equations for the species densities and species velocities,
\begin{align}
\partial_t\rho_a&=-\nabla\cdot(\rho_a \bm{u}_a),
\label{eq:dtrhoa}\\
\rho_a\partial_t \bm{u}_a&=\bm{u}_a\nabla\cdot(\rho_a\bm{u}_a)-\nabla\cdot \bm{P}_a+\sum_{b=1}^M \frac{1}{\theta_{ab}} \left( \bm{u}_{b} - \bm{u}_a \right).
\label{eq:dtua}
\end{align}
Here, $\bm{P}_a$ is the partial pressure tensor,
\begin{equation}
\bm{P}_a=\sum_{i=0}^{Q-1} f_{ai}\bm{c}_i\otimes \bm{c}_i.
\end{equation}
Upon summation over the components in (\ref{eq:dtrhoa}), we arrive at the continuity equation for the mixture density,
\begin{equation}
    \partial_t\rho=-\nabla\cdot(\rho\bm{u}),
    \label{eq:continuitySPEC}
\end{equation}
while the summation over components in (\ref{eq:dtua}) results in the mixture momentum balance,
\begin{equation}
    \partial_t(\rho\bm{u})=-\nabla\cdot\bm{P},
    \label{eq:momentumSPEC}
\end{equation}
where $\bm{P}$ is the mixture pressure tensor,
\begin{equation}
    \bm{P}=\sum_{a=1}^{M}\bm{P}_a.
\end{equation}

The low-order closure relation for the species balance equation (\ref{eq:dtrhoa}) is established by considering a perturbation around the equilibrium,
\begin{equation}
\bm{u}_a=\bm{u}+\delta\bm{u}_a,
\label{eq:perturb}
\end{equation}
where the perturbation $\delta\bm{u}_a$ satisfies the consistency condition,
\begin{equation}
    \sum_{a=1}^{M}\rho_a\delta\bm{u}_a=0.
    \label{eq:consistent}
\end{equation}
To first order, upon substitution into (\ref{eq:dtua}), we get the constitutive relation for the diffusion velocity $\bm{u}_a$,
\begin{equation}
\rho_a\partial_t\bm{u}-\bm{u}\nabla\cdot(\rho_a\bm{u})+\nabla\cdot\bm{P}_a^{\rm eq}=\sum_{b=1}^M \frac{1}{\theta_{ab}} \left( \bm{u}_{b} - \bm{u}_a \right).
\label{eq:constit}
\end{equation}
Upon summation over the species, and by taking into account Dalton's law in the equilibrium pressure tensor (\ref{eq:Pa}), the compressible Euler equation for the flow velocity is established,
\begin{equation}
\partial_t\bm{u}=-\bm{u}\cdot\nabla\bm{u}-\frac{1}{\rho}\nabla{P}.
\end{equation}
By elimination of the time derivative in (\ref{eq:constit}), we get the constitutive relation as follows:
\begin{equation}
P\nabla X_a+(X_a-Y_a)\nabla P=\sum_{b=1}^M \frac{1}{\theta_{ab}} \left( \bm{u}_{b} - \bm{u}_a \right).
\label{eq:constit2}
\end{equation}
The constitutive relation (\ref{eq:constit2}) becomes the Stefan--Maxwell diffusion equation once the relaxation parameters $\theta_{ab}$ are identified in terms of the binary mass diffusion coefficients $D_{ab}$ as follows:
%
%
\begin{equation}
\theta_{ab}=\frac{D_{ab}}{PX_aX_b}.
\label{eqn:smDiffusionCoefficients2}
\end{equation}
Summarizing, kinetic equations for the species (\ref{eqn:stefanMaxwell}) recover the Stefan--Maxwell law of diffusion in the hydrodynamic limit, with both the diffusion due to non-uniformity of the species concentration and the barodiffusion taken into account. The present model does not include thermodiffusion as it should be expected by the simplicity of the relaxation term. We comment that the above derivation of (\ref{eq:constit2}) assumes the validity of the equation of state. The latter, in turn, depends on the temperature derived from the mixture energy equation, and which shall be introduced in Sec.\ \ref{sec:lbMixtures}. 
We now proceed with finalizing the continuous time-space kinetic equations by identifying the equilibrium and the quasi-equilibrium populations.

\subsection{Realization on the standard lattice}
\label{sec:27spec}

The above kinetic model is realized on the standard three-dimensional $D3Q27$ lattice, where $D=3$ stands for three dimensions and $Q=27$ is the number of discrete velocities:
\begin{equation}\label{eq:d3q27vel}
	\bm{c}_i=(c_{ix},c_{iy},c_{iz}),\ c_{i\alpha}\in\{-1,0,1\}.
\end{equation}
Following \citet{karlin2010factorization}, we define a triplet of functions in two variables, $\xi$ and $\zeta>0$,
\begin{align}
	\Psi_{0}(\xi,\zeta) = 1 - (\xi^2 + \zeta), \
	\Psi_{1}(\xi,\zeta) = \frac{\xi + (\xi^2 + \zeta)}{2},\ 
	\Psi_{-1}(\xi,\zeta) = \frac{-\xi + (\xi^2 + \zeta)}{2}.
\end{align}
For a vector $\bm{\xi}=(\xi_x,\xi_y,\xi_z)$, we consider a product-form associated with the discrete velocities $\bm{c}_i$ (\ref{eq:d3q27vel}),
\begin{equation}\label{eq:prod}
	\Psi_i(\bm{\xi},\zeta)= \Psi_{c_{ix}}(\xi_x,\zeta) \Psi_{c_{iy}}(\xi_y,\zeta) \Psi_{c_{iz}}(\xi_z,\zeta).
\end{equation}
%
%
%
The equilibrium $f_{ai}^{\rm eq}$ and the quasi-equilibrium $f_{ai}^*$ are represented with the product-form (\ref{eq:prod}) by choosing $\bm{\xi}=\bm{u}$ or $\bm{\xi}=\bm{u}_a$, respectively, and by assigning $\zeta=R_aT$ in both cases:
%
%
%
\begin{align}
	f_{ai}^{\rm eq}(\rho_a,\bm{u},T)&= \rho_a\Psi_{c_{ix}}\left(u_x,R_aT\right) \Psi_{c_{iy}}\left(u_y,R_aT\right) \Psi_{c_{iz}}\left(u_z,R_aT\right),
\label{eq:27eq}	\\
	f_{ai}^{*}(\rho_a,\bm{u}_a,T)&= \rho_a\Psi_{c_{ix}}\left(u_{ax},R_aT\right)
	\Psi_{c_{iy}}\left(u_{ay},R_aT\right) 
	\Psi_{c_{iz}}\left(u_{az},R_aT\right).
\label{eq:27qeq}
\end{align}
One can readily verify that, the equilibrium (\ref{eq:27eq}) and the quasi-equilibrium (\ref{eq:27qeq}) satisfy all the constraints put forward in sec.\ \ref{sec:formulation}.
We now proceed with the lattice Boltzmann discretization of the kinetic equations (\ref{eqn:stefanMaxwell}).

\subsection{Lattice Boltzmann equation for the species}
\label{sec:numericalImplementation}

\subsubsection{Kinetic equations in the relaxation form}
With the mass diffusivity (\ref{eqn:smDiffusionCoefficients2}), the kinetic equation (\ref{eqn:stefanMaxwell}) is written as
\begin{equation}
\partial_t f_{ai} + \bm{c}_{i}\cdot\nabla f_{ai} = \sum_{b=1}^M \frac{P X_a X_b}{D_{ab}} \left[ \left( \frac{f_{ai}^{\rm eq}-f_{ai}}{\rho_a} \right) - \left( \frac{f_{bi}^{\rm eq}-f^*_{bi}}{\rho_b} \right) \right].
\label{eqn:stefanMaxwellNumerical} 
\end{equation}
It can readily be seen that, in its present form, Eq.\ (\ref{eqn:stefanMaxwellNumerical}) is  not well suited for  numerical implementation. Indeed, in an actual problem, the density of some species can be small or even vanishing if a particular gas component is absent at some location. This is an inconvenience rather than a failure since vanishing of the density is compensated by the simultaneously vanishing molar fraction in the product $X_aX_b$. 
Hence, we first transform equation (\ref{eqn:stefanMaxwellNumerical}) in order to eliminae this  artifact.  
Substituting the equation of state (\ref{eqn:eosIdealGas}) into Eq.\ (\ref{eqn:stefanMaxwellNumerical}), we get
\begin{align}
\partial_t f_{ai} + \bm{c}_{i}\cdot\nabla f_{ai} &= 
\sum_{b=1}^M \left(\frac{  m }{m_a m_b}\right)\left(\frac{ R_UT }{D_{ab}}\right) 
\left[ Y_b \left( f_{ai}^{\rm eq}-f_{ai} \right) - Y_a \left( f_{bi}^{\rm eq}-f^*_{bi} \right) \right].
\label{eqn:stefanMaxwellNumericalStable} 
\end{align}
Equation (\ref{eqn:stefanMaxwellNumericalStable}) is equivalent to equation (\ref{eqn:stefanMaxwellNumerical}) and does not suffer from a spurious division by a vanishing density. 
Furthermore, it proves convenient to recast Eq.\ (\ref{eqn:stefanMaxwellNumericalStable}) in a relaxation form. To that end, let us define  characteristic times $\tau_{ab}=\tau_{ba}$,
\begin{equation}
\frac{1}{\tau_{ab}} = \left(\frac{R_UT}{D_{ab}}\right)\left(\frac{m}{m_a m_b}\right),
\label{eqn:tauab}
\end{equation}
and let us introduce the relaxation times $\tau_a$,
\begin{equation}
\frac{1}{\tau_a} = \sum_{b=1}^M \frac{Y_b}{\tau_{ab}} 
\label{eqn:tau}
\end{equation}
Finally, let us introduce a shorthand notation,
\begin{equation}
F_{ai} = Y_a \sum_{b=1}^M \frac{1}{\tau_{ab}}  \left( f_{bi}^{\rm eq}-f_{bi}^* \right).
\label{eqn:fStar}
\end{equation}
With these definitions, the kinetic equation (\ref{eqn:stefanMaxwellNumericalStable}) can be rearranged as follows: 
\begin{equation}
\partial_t f_{ai} + \bm{c}_{i}\cdot\nabla f_{ai}  =  \frac{1}{\tau_a}\left( f_{ai}^{\rm eq}-f_{ai} \right)  - F_{ai}. 
\label{eqn:cleanEquation} 
\end{equation}
The species kinetic equations (\ref{eqn:cleanEquation}) are now cast into the relaxation form, familiar from previous lattice Boltzmann models: The right hand side comprises the conventional relaxation term and a source term (\ref{eqn:fStar}). The latter depends on the populations only through the local densities,  momenta and the temperature, as prescribed by the local equilibrium and quasi-equilibrium populations (\ref{eq:27eq}) and (\ref{eq:27qeq}). Hence, kinetic equation (\ref{eqn:cleanEquation}) is amenable to a lattice Boltzmann discretization in time and space.

\subsubsection{Derivation of the lattice Boltzmann equation}

Following a procedure first introduced by \cite{he1998}, we integrate (\ref{eqn:cleanEquation}) along the characteristics and apply the trapezoidal rule on the right hand side to obtain,
\begin{multline}
f_{ai}(\bm{x}+\bm{c}_i \delta t, t+ \delta t) - f_{ai}(\bm{x},t) =
 \frac{\delta t}{2 \tau_a}[f_{ai}^{\rm eq}(\bm{x}+\bm{c}_i \delta t, t+ \delta t) - f_{ai}(\bm{x}+\bm{c}_i \delta t, t+ \delta t)] \\
+ \frac{\delta t}{2 \tau_a}[f_{ai}^{\rm eq}(\bm{x},t) - f_{ai}(\bm{x},t)] - \frac{\delta t}{2} F_{ai}(\bm{x}+\bm{c}_i \delta t, t+ \delta t) 
- \frac{\delta t}{2} F_{ai}(\bm{x}, t).
\label{eqn:trapezoidalRule}
\end{multline}
Next, we introduce transformed  populations $k_{ai}$,
\begin{equation}
f_{ai} = k_{ai} + \frac{\delta t}{2 \tau_a} (f_{ai}^{\rm eq}-f_{ai}) - \frac{\delta t}{2} F_{ai}.
\label{eqn:transform}
\end{equation}
Let us evaluate the pertinent moments of the transform (\ref{eqn:transform}). Summation over the discrete velocities gives,
\begin{align}
	\rho_a(f) = \rho_a(k),
	\label{eqn:transform0}
\end{align}
where we have specified that the density $\rho_a(f)$ on the left hand side is defined using the original populations $f_{ai}$ while the density $\rho_a(k)$ is defined by the zeroth moment of the $k$-populations,
\begin{equation}
\rho_a(k)=\sum_{i=0}^{Q-1}k_{ai}.
\end{equation}
Thus, the species densities do not alter under the populations transformation, and the specification can be dropped: $\rho_a=\rho_a(f)=\rho_a(k)$.
 On the other hand, evaluating the first moment of (\ref{eqn:transform}) gives,
\begin{align}
\rho_a \bm{u}_{a}(f) \left( 1+ \frac{\delta t}{2 \tau_a}\right) - \frac{\delta t}{2} Y_a \sum_{b=1}^{M} \frac{1}{\tau_{ab}} \rho_b\bm{u}_{b }(f)=\rho_a \bm{u}_{a}(k),
\label{eqn:transform1}
\end{align}
where the species velocity $\bm{u}_a(k)$ is defined by the $k$-populations in a conventional way,
\begin{equation}
\rho_a\bm{u}_a(k)=\sum_{i=0}^{Q-1}k_{ai}\bm{c}_i.
\end{equation}
Summation over the species in (\ref{eqn:transform1}) shows that the momentum $\rho\bm{u}$ is also an invariant of the transform (\ref{eqn:transform}):
\begin{equation}
    \rho\bm{u}(f)=\rho\bm{u}(k).
    \label{eq:transformU}
\end{equation}
Since the term $F_{ai}$ vanishes at equilibrium, and also thanks to the invariance of the local conservation (\ref{eqn:transform0}) and (\ref{eq:transformU}), the equilibrium is the fixed point of the map (\ref{eqn:transform}):
\begin{equation}\label{eq:fixed}
f_{ai}^{\rm eq}(\rho_a,\bm{u},T)=k_{ai}^{\rm eq}(\rho,\bm{u},T).
\end{equation}

%
%
%
Substituting (\ref{eqn:transform}) into Eq.\ (\ref{eqn:trapezoidalRule}), and introducing the parameters $\beta_a\in[0,1]$,
%
%
\begin{equation}
\beta_a=\frac{\delta t}{2 \tau_a + \delta t},
\end{equation} 
we get,
\begin{equation}
 k_{ai}(\bm{x}+\bm{c}_i \delta t, t+ \delta t)  = k_{ai}(\bm{x},t)+ 2 \beta_a [k_{ai}^{\rm eq}(\bm{x},t) - k_{ai}(\bm{x},t)]
+ \delta t (\beta_a-1) F_{ai}(\bm{x}, t).
\label{eqn:finalNumericalEquations}
\end{equation}
The last term in Eq.\ (\ref{eqn:finalNumericalEquations}) is spelled out as follows: The quasi-equilibrium population $f_{ai}^*$ in
the expression $F_{ai}$ (\ref{eqn:fStar}) depends on the species velocity $\bm{u}_a(f)$. The latter, unlike the mixture velocity $\bm{u}(f)$ 
(\ref{eq:transformU}), is not invariant under the transform to the $k$-populations. Rather, $\bm{u}_a(f)$ has to be evaluated using the linear relation (\ref{eqn:transform1}) in terms of $\bm{u}_b(k)$ by solving a $M\times M$ linear system for each of the spatial components.

\subsubsection{Summary of the lattice Boltzmann equation for the Stefan--Maxwell diffusion}
\label{sec:LBMspec}

For convenience, we summarize the lattice Boltzmann equation for the species. 
We return to a more conventional notation and rename $k_{ai}$ to $f_{ai}$ in (\ref{eqn:finalNumericalEquations}), 
\begin{align}
 f_{ai}(\bm{x}+\bm{c}_i \delta t, t+ \delta t) - f_{ai}(\bm{x},t)= 2 \beta_a [f_{ai}^{\rm eq} - f_{ai}]
+ \delta t (\beta_a-1) F_{ai},
\label{eqn:SMlbm}
\end{align}
The last term is written 
\begin{align}
&F_{ai} = Y_a \sum_{b=1}^M \left(\frac{R_UT}{D_{ab}}\right)\left(\frac{m}{m_a m_b}\right)
\left[ f_{bi}^{\rm eq}(\rho_b,\bm{u},T)-f_{bi}^*(\rho_b,\bm{V}_b+\bm{u},T) \right],
\label{eq:fstar1}
\end{align}
where we have introduced the diffusion velocity of the components, $\bm{V}_a$, $a=1,\dots,M$. The latter are determined by 
 (\ref{eqn:transform1}) which can be recast as follows: 
%
\begin{equation}\label{eq:transform12}
	\rho_a \bm{V}_{a}-\frac{\delta t}{2}\sum_{b=1}^{M}\frac{PX_aX_b}{D_{ab}}[\bm{V}_{b }-\bm{V}_{a }]=\rho_a \bm{u}_{a}-\rho_a\bm{u}.
\end{equation}
The field $\rho_a\bm{u}_a$ in the right hand side of  (\ref{eq:transform12}) is defined by the moment relation as before,
\begin{align}\label{eq:auxua}
    \rho_a\bm{u}_a=\sum_{i=0}^{Q-1}f_{ai}\bm{c}_i.
\end{align}
The $M+D$ independent conservation laws of the $M$-component system (\ref{eqn:SMlbm}) correspond to the mass of each component and the momentum of the mixture. The corresponding locally conserved fields are the species densities $\rho_a$ and the momentum flux $\rho\bm{u}$,
\begin{align}
    \rho_a&=\sum_{i=0}^{Q-1}f_{ai},\label{eq:densitya}\\
    \rho\bm{u}&=\sum_{a=1}^M \rho_a\bm{u}_a=\sum_{a=1}^M\sum_{i=0}^{Q-1}f_{ai}\bm{c}_i.\label{eq:momSPEC}
\end{align}
The conservation law of the fluid mass is the implication of the mass conservation of each component. 
The density of the mixture $\rho$ is defined by the densities of the components,
\begin{align}
    \label{eq:rhoSPEC}
    \rho&=\sum_{a=1}^M \rho_a=\sum_{a=1}^M\sum_{i=0}^{Q-1}f_{ai},
\end{align}
while the flow velocity $\bm{u}$ is,
\begin{align}
    \label{eq:uSPEC}
    \bm{u}&=\frac{\rho\bm{u}}{\rho}=\frac{\sum_{a=1}^M\sum_{i=0}^{Q-1}f_{ai}\bm{c}_i}{\sum_{a=1}^M\sum_{i=0}^{Q-1}f_{ai}}.
\end{align}
Note that, the velocity of the component is defined as the sum of the diffusion velocity and the flow velocity, $\bm{V}_a+\bm{u}$, in
the quasi-equilibrium populations 
contributing in (\ref{eq:fstar1}), rather than as the first moment of the populations (\ref{eq:auxua}).
Finally, we remind that the temperature dependence in the equilibrium and the quasi-equilibrium populations has to be supplied by the energy equation to be discussed in the next section.

\section{Lattice Boltzmann model of mixture momentum and energy}
\label{sec:lbMixtures}

\subsection{First law of thermodynamics for ideal gas mixture}
\label{sec:thermo2}
For further references and notation, we open this section with a summary of the first law of thermodynamics for ideal gas mixtures.
The caloric equation of state of a single-component ideal gas provides for the specific mole-based internal energy of species $a$:
\begin{align}\label{eq:eoscal}
\bar{U}_a=\int_{T_0}^T\bar{C}_{a,v}(T)dT.
\end{align}
Here, $\bar{C}_{a,v}$ is the specific heat at constant volume. Thus, the specific enthalpy reads,
\begin{align}
\bar{H}_a=\int_{T_0}^T\bar{C}_{a,p}(T)dT,
\end{align}
where $\bar{C}_{a,p}$ is the specific heat at constant pressure defined by Mayer's relation,
\begin{equation}
\bar{C}_{a,p}-\bar{C}_{a,v}=R_{U}.
\end{equation}
Proceeding from the mole-basis onto the mass-basis, the specific heats are defined relative to the molar mass,
\begin{align}
{C}_{a,v}=\frac{\bar{C}_{a,v}}{m_a},\\
{C}_{a,p}=\frac{\bar{C}_{a,p}}{m_a},
\end{align}
while the mass-based specific internal energy and enthalpy are,
\begin{align}
{U}_a&=\int_{T_0}^T{C}_{a,v}(T)dT,
\label{eq:specUa}\\
{H}_a&=\int_{T_0}^T{C}_{a,p}(T)dT.
\label{eq:specHa}
\end{align}
Finally, the Mayer relation in the mass-basis reads,
\begin{equation}
{C}_{a,p}-{C}_{a,v}=R_{a},
\end{equation}
where the gas constant $R_a$ is defined by (\ref{eq:Ra}).

Switching to the case of a $M$-component mixture, the mixture internal energy $\rho U$ is defined on the  mass-basis as follows:
\begin{equation}
\rho U=\sum_{a=1}^M\rho_a U_a.
\label{eq:U}
\end{equation}
The specific mixture internal energy $U$ can be rewritten,
\begin{equation}
 U=\sum_{a=1}^MY_a U_a=\sum_{a=1}^MY_a \int_{T_0}^T{C}_{a,v}dT=\int_{T_0}^T\left[\sum_{a=1}^MY_a{C}_{a,v}\right]dT=\int_{T_0}^T{C}_{v}dT,
 \label{eq:Umix}
\end{equation}
where the specific heat at constant volume is the mass-averaged value over the composition,
\begin{equation}\label{eq:Cvmix}
{C}_{v}=\sum_{a=1}^MY_a{C}_{a,v}
\end{equation}
Similarly, the mixture enthalpy $\rho H$ is defined as,
\begin{equation}
\rho H=\sum_{a=1}^M\rho_a H_a,
\end{equation}
while the specific mixture enthalpy $H$ can be transformed in the manner of Eq.\ (\ref{eq:Umix}),
\begin{equation}
H=\sum_{a=1}^MY_a H_a=\sum_{a=1}^MY_a \int_{T_0}^T{C}_{a,p}dT=\int_{T_0}^T\left[\sum_{a=1}^MY_a{C}_{a,p}\right]dT=\int_{T_0}^T{C}_{p}dT.
\label{eq:specH}
\end{equation}
The specific heat at constant pressure reads,
\begin{equation}
{C}_{p}=\sum_{a=1}^MY_a{C}_{a,p},
\end{equation}
while both the specific heats satisfy the Mayer relation,
\begin{equation}
{C}_{p}-{C}_{v}=R,
\end{equation}
with the mixture gas constant $R$ defined by Eq.\ (\ref{eqn:defineSpecificGasContant}). Below, we shall formulate the lattice Boltzmann equation for the mixture density, momentum and energy for a generic case of temperature-dependent specific heats of the components.

\subsection{Two-population lattice Boltzmann equation for the mixture}
\label{sec:kinetic}

Point of departure is a lattice Boltzmann model for a {\it single}-component ideal gas with variable Prandtl number and adiabatic exponent. To that end, several suitable single-component lattice Boltzmann models exist in the literature; here we mention compressible LBM by \cite{frapolli2015,frapolli2016entropic} and by \citet{hosseinCompressible}. The common feature of these single-component models is the use of the {\it double-population construction}, the idea first introduced in the context of incompressible thermal convective LBM in the classical paper by \cite{he1998} and further expanded in 
\cite{guo2007twopop,karlinConsistent,frapolli2018}.   One set of populations, commonly quoted as $f$-populations, represents the density and momentum as the locally conserved fields of the corresponding $f$-LBM equation while another set, the $g$-populations represent the energy as the local conservation of the $g$-LBM kinetics. The coupling between the $f$- and $g$-LBM equations is also well understood and enables the realization of an adjustable Prandtl number and adiabatic exponent. Various realizations differ by the choice of the discrete velocities of the $f$- and $g$-sets; in particular, the compressible LBM of \cite{frapolli2015} employs higher-order lattices with higher isotropy while the two-dimensional model developed in \citet{hosseinCompressible} uses the standard lattice with correction terms to compensate for insufficient isotropy.  

Whichever {\it single}-component double-population model is taken as the starting point for representing a {\it multi}-component mixture, the central question is how to modify it. Note that, this question would not arise if one would follow the conventional approach by extending the already available $M$ species LBM equations of sec.\ \ref{sec:stefanMaxwell} to represent the energy equation of the mixture. However, with the double-population approach, this would lead to $2\times M$ lattices since the lattice for each component would need to be doubled to represent the energy of that component. On the contrary, the {\it mean-field} approach pursuit here avoids the kinetic representation of partial energies, instead it addresses only the total energy of the mixture by a single $g$-set. This requires only $M+2$ lattices, $M$ for the species and two for the mixture momentum and energy.

Below, we refer to the $f$-populations as the momentum lattice, and the $g$-populations as the energy lattice.
For the momentum lattice, the locally conserved fields are the density and the momentum of the mixture,
\begin{align}
&\sum_{i=0}^{Q-1} f_i  = \rho,
\label{eqn:f0momDensity} 
\\
&\sum_{i=0}^{Q-1} f_i \bm{c}_{i} = \rho \bm{u}.
\label{eqn:f1momMomentum} 
\end{align}
For the energy lattice, the locally conserved field is the total energy of the mixture,
\begin{align}
&\sum_{i=0}^{Q-1} g_i  =  \rho E. 
\label{eqn:g0momTotalEnergy}
\end{align}
Here, the total energy $\rho E$ is the sum of the mixture internal energy $\rho U$ (\ref{eq:Umix}) and the kinetic energy $\rho u^2/2$, 
\begin{align}
\rho E=\rho U + {\frac{\rho u^2}{2}}.
\end{align}
Since the mixture internal energy (\ref{eq:Umix}) depends on the composition, it is the first instance where the species kinetic equations become coupled with the kinetic equations for the mixture. Conversely, the temperature of the mixture is computed from the  integral equation,

\begin{equation}
\label{eq:temperature}
    \int_{T_0}^T C_{v}(T)dT=E-\frac{u^2}{2}.
\end{equation}
In the simplest case, the specific heats of all components can be approximated by constants and  the temperature becomes
\begin{equation}
\label{eq:temperature2}
    T=T_0+\frac{E-(u^2/2)}{\sum_{a=1}^M Y_aC_{a,v}}.
\end{equation}
The temperature evaluated from solving (\ref{eq:temperature}) is used as the input in the definition of the pressure $P$ in the equilibrium and quasi-equilibrium constraints of the species lattice Boltzmann system. This furnishes the input from the energy lattice into the species lattices.

We comment that, in the present section, the mixture density (\ref{eqn:f0momDensity}) and momentum (\ref{eqn:f1momMomentum}) are defined self-consistently in the sense of $f$-populations of the momentum lattice. On the other hand, quantities carrying the same physical meaning were independently and also self-consistently defined earlier using the species populations, Eqs.\ (\ref{eq:rhoSPEC}) and (\ref{eq:momSPEC}), respectively. Doubling of the conservation of the total mass and momentum  is the feature of the intermediate steps of the construction during which the species subsystem and the mixture subsystem  are set up independently from one another. At the end of the construction, the doubling of the conservation shall be resolved through a coupling of both the species and the mixture subsystems in Sec.\ \ref{sec:coupling}. 





The lattice Boltzmann equations for the momentum and for the energy lattice are patterned from the single-component developments,
\begin{align}
f_i(\bm{x}+\bm{c}_i \delta t,t+\delta t)- f_i(\bm{x},t)&=  \omega (f_i^{\rm eq} -f_i),  \label{eqn:f} 
\\
g_i(\bm{x}+\bm{c}_i \delta t,t+ \delta t) - g_i(\bm{x},t)&=  \omega_1 (g_i^{\rm eq} -g_i) + (\omega - \omega_1) (g_i^{*} -g_i),
 \label{eqn:g}
\end{align}
where relaxation parameters $\omega$ and $\omega_1$ shall be related to the viscosity and thermal conductivity below.
We now proceed with specifying the constraints on the equilibrium populations $f_i^{\rm eq}$ and $g_i^{\rm eq}$, and the quasi-equilibrium $g_i^{*}$ in order that the system (\ref{eqn:f}) and (\ref{eqn:g}) recovers the momentum and energy equations of the mixture.

First, the equilibrium populations must satisfy the  $D+2$ conservation laws,
\begin{align}
& \sum_{i=0}^{Q-1} f_i^{\rm eq} = \rho,
\label{eqn:f0mom} 
\\
&\sum_{i=0}^{Q-1} f_i^{\rm eq} \bm{c}_{i } = \rho \bm{u}, 
\label{eqn:f1mom} 
\\
& \sum_{i=0}^{Q-1} g_i^{\rm eq}  =  \rho E. 
\label{eqn:g0mom}
\end{align}
Second, the equilibrium pressure tensor $\bm{P}^{\rm eq}$ and the tensor of equilibrium third-order moments $\bm{Q}^{\rm eq}$ of the momentum lattice must verify the Maxwell--Boltzmann relations in order to recover the compressible flow momentum equation,
\begin{align}
&\bm{P}^{\rm eq} =\sum_{i=0}^{Q-1} f_i^{\rm eq} \bm{c}_i\otimes\bm{c}_i = P \bm{I}+\rho \bm{u}\otimes\bm{u},
\label{eqn:feq2mom} 
\\
&\bm{Q}^{\rm eq} =\sum_{i=0}^{Q-1} f_i^{\rm eq} \bm{c}_i\otimes\bm{c}_i\otimes\bm{c}_i = P\overline{\bm{u}\otimes\bm{I}} + \rho \bm{u}\otimes\bm{u}\otimes\bm{u},
\label{eqn:feq3mom} 
\end{align}
where overline denotes symmetrization. Similarly, the equilibrium mixture energy flux  $\bm{q}^{\rm eq}$ and the second-order moment tensor $\bm{R}^{\rm eq}$ pertinent to the energy lattice are,
\begin{align}
&\bm{q}^{\rm eq}= \sum_{i=0}^{Q-1}  g_i^{\rm eq} \bm{c}_{i} =  \left(H+\frac{u^2}{2}\right)\rho\bm{u},
\label{eqn:geq1mom} 
\\
&\bm{R}^{\rm eq}=\sum_{i=0}^{Q-1} g_i^{\rm eq} \bm{c}_i\otimes\bm{c}_i =
    \left(H+\frac{u^2}{2}\right) \bm{P}^{\rm eq} + P\bm{u}\otimes\bm{u}.
\label{eqn:geq2mom}
\end{align}
The mixture equation of state $P$ (\ref{eqn:eosIdealGas}), the mixture gas constant $R$ (\ref{eqn:defineSpecificGasContant}) and the specific enthalpy of the mixture $H$ (\ref{eq:specH}) entering the constraints (\ref{eqn:feq2mom}), (\ref{eqn:feq3mom}), (\ref{eqn:geq1mom}), (\ref{eqn:geq1mom}) and (\ref{eqn:geq2mom})  depend linearly on the composition through the local mass fractions $Y_a$. 

To that end, the constraints on the equilibrium populations of the mixture momentum and energy lattices is a straightforward extension of those of the single-component double-population LBM for compressible flow, where the ideal gas equation of state, the internal energy and the enthalpy are merely replaced by their mixture-averaged counterparts. A major difference comes next with the constraints for the quasi-equilibrium. The zeroth, the first- and the second-order moments of the quasi-equilibrium populations $g_i^{*}$, or the quasi-equilibrium energy $\rho E^{*}$, the energy flux $\bm{q}^{*}$ and the flux of the energy flux $\bm{R}^{*}$, respectively, have to satisfy the following relations:
\begin{align}
&\rho E^{*}=\sum_{i=0}^{Q-1} g_i^{*}=\rho E, \label{eqn:gstareq0mom}\\
&\bm{q}^{*} =\sum_{i=0}^{Q-1} g_i^{*} \bm{c}_{i} =  \bm{q} -  \bm{u}\cdot (\bm{P} - \bm{P}^{\rm eq}) +\bm{q}^{\rm diff}+\bm{q}^{\rm corr}
\label{eqn:gstareq1mom} \\
&\bm{R}^{*}=\sum_{i=0}^{Q-1} g_i^{*} \bm{c}_{i}\otimes \bm{c}_i=\bm{R}^{\rm eq}.\label{eqn:gstareq2mom}
\end{align}
The first and the third of these quasi-equilibrium constraints, Eqs.\ (\ref{eqn:gstareq0mom}) and (\ref{eqn:gstareq2mom}), as well as the first and the second terms in the quasi-equilibrium energy flux (\ref{eqn:gstareq1mom}) are again the direct extension of the single-component LBM. Specifically, the two first terms in (\ref{eqn:gstareq1mom}), comprising the energy flux $\bm{q}$ and the pressure tensor $\bm{P}$,
\begin{align}
    \bm{q}&=\sum_{i=0}^{Q-1} g_i \bm{c}_{i},\label{eq:q}\\
    \bm{P}&= \sum_{i=0}^{Q-1} f_i \bm{c}_{i}\otimes \bm{c}_{i}, \label{eq:boldP}
\end{align}
are needed to decouple the viscosity from thermal conductivity, and to maintain a variable Prandtl number, in both the single- and multcomponent cases.

The remaining two terms in the quasi-equilibrium energy flux (\ref{eqn:gstareq1mom}), $\bm{q}^{\rm diff}$ and $\bm{q}^{\rm corr}$ are specific to the multicomponent case and appear due to the mean-field approach to the energy representation.
%
%
%
%
The interdiffusion energy flux $\bm{q}^{\rm diff}$ reads as follows:
\begin{align}
	\bm{q}^{\rm diff}=\left(\frac{\omega_1}{\omega-\omega_1} \right) \rho\sum_{a=1}^{M}H_aY_a\bm{V}_a,
	\label{eq:interdiffusion}
\end{align}
where the diffusion velocities $\bm{V}_a$ are defined according to Eq.\ (\ref{eq:transform12}). 
%
The interdiffusion energy flux contributes the enthalpy transport due to diffusion and hence it vanishes in the single-component case. The effect of the inter-diffusion energy flux is typically significant at the initial stages of the diffusion process and cannot be neglected.

Finally, the correction flux $\bm{q}^{\rm corr}$ reads, 
\begin{align}
\bm{q}^{\rm corr}=\frac{1}{2}\left(\frac{\omega_1-2}{\omega_1-\omega}\right) {\delta t}P \sum_{a=1}^M  H_{a}\nabla Y_a,
\label{eq:corrFourier}
\end{align}
and is explained by the following consideration: The thermal flux is the mixture-average of the component  thermal fluxes,
$\bm{q}^{\rm th}=\sum_{a=1}^{M}Y_a \bm{q}_a^{\rm th}$, where $\bm{q}_a^{\rm th}= -\tau PC_{a,p}\nabla T$ is the Fourier law for the component, $\tau$ is a parameter of no importance to the current consideration. On the other hand, in the single-component LBM, the thermal flux is $\bm{q}^{\rm th}_{\rm sc}= - \tau P\nabla H_{sc}$, and with the single-component enthalpy $H_{\rm sc}$ it returns the Fourier law in this case. However, the extension of the single- to the multicomponent case so far invokes only the replacement of the single-component enthalpy with the ``lumped" mixture enthalpy and without any correction one gets,
\[\bm{q}^{\rm lump}= -\tau P\nabla\left(\sum_{a=1}^{M}Y_aH_a\right)=\bm{q}^{\rm th}-\tau P\sum_{a=1}^{M}H_a\nabla Y_a.\]
Thus, apart from the mixture-averaged Fourier law $\bm{q}^{\rm th}$, the thermal flux also contains a spurious term. 
The spurious term is eliminated by the correction flux $\bm{q}^{\rm corr}$ (\ref{eq:corrFourier}),  where the prefactor is chosen by considering the 
hydrodynamic limit;
see Appendix \ref{sec:ceNavierStokes}. 
The correction flux vanishes if all components are thermodynamically indistinguishable, that is, if all species have the same specific heat. In many cases, the correction flux contributes negligibly, for example, for air at moderate temperatures where the standard-air assumptions for diatomic molecules holds to a good approximation.

\subsection{Hydrodynamic limit of the two-population lattice Boltzmann model for mixtures}
\label{sec:hydromixture}

Constraints on the pertinent equilibrium and quasi-equilibrium moments
(\ref{eqn:feq2mom}), (\ref{eqn:feq3mom}),  
(\ref{eqn:geq1mom}), (\ref{eqn:geq2mom}),  
(\ref{eqn:gstareq0mom}), (\ref{eqn:gstareq1mom}), (\ref{eq:interdiffusion}), (\ref{eq:corrFourier}) and (\ref{eqn:gstareq2mom})
are sufficient to study the hydrodynamic limit of the two-population lattice Boltzmann system (\ref{eqn:f}) and (\ref{eqn:g}) without a complete specification of the equilibrium and the quasi-equilibrium populations. The analysis follows the route of expanding the propagation to second order in the time step $\delta t$ and evaluating the moments of the resulting expansion.  Details of the derivation are included in Appendix \ref{sec:ceNavierStokes}, here we present the final result:
%

\noindent The continuity equation:
\begin{equation}
\partial_t \rho + \nabla\cdot (\rho \bm{u})=0.
\label{eqn:dtrho}
\end{equation}
The momentum equation:
\begin{equation}
\partial_t (\rho\bm{u}) +  \nabla\cdot ({\rho\bm{u}\otimes\bm{u} })+ \nabla\cdot \bm{\pi}=0.
\label{eqn:dtu}
\end{equation}
Here, the pressure tensor $\bm{\pi}$ reads,
\begin{equation}
\bm{\pi}=P\bm{I}
-\mu \left( \bm{S}  -\frac{2}{D} (\nabla\cdot\bm{u})\bm{I} \right) 
-\varsigma (\nabla\cdot\bm{u}) \bm{I},
\end{equation}
where $\bm{S}$ is the strain rate,
\begin{equation}
\bm{S}=\nabla\bm{u}  + \nabla\bm{u}^{\dagger}.
\end{equation}
The dynamic viscosity $\mu$ and the bulk viscosity $\varsigma$ are related to the relaxation parameter $\omega$,
\begin{align}
\mu & = \left( \frac{1}{\omega} - \frac{1}{2}\right) P{\delta t},
\label{eq:mu}\\
\varsigma &=\left( \frac{1}{\omega}-\frac{1}{2}\right)\left( \frac{2}{D} - \frac{R}{C_v} \right) P{\delta t}.
\label{eqn:varsigma}
\end{align}

\noindent The energy equation:
\begin{align}
\partial_t (\rho E)+\nabla\cdot(\rho E\bm{u})+\nabla\cdot\bm{q}+\nabla\cdot(\bm{\pi}\cdot\bm{u})=0.
\end{align}
Here, the heat flux $\bm{q}$ reads,
\begin{equation}
\label{eq:qneq}
\bm{q}=-\lambda\nabla T+\rho\sum_{a=1}^{M}H_aY_a\bm{V}_a.
\end{equation}
The first term is the Fourier law of thermal conduction, with thermal conductivity $\lambda$ related to the relaxation parameter $\omega_1$,
\begin{equation}\label{eq:lambda}
\lambda= \left(\frac{1}{\omega_1} - \frac{1}{2}\right) P C_p{\delta t}.
\end{equation}
The second term in (\ref{eq:qneq}) is the interdiffusion energy flux. Some comments are in order:
\begin{itemize}
    \item The continuity, the momentum and the energy equations are the standard equations for a multicomponent compressible mixtures \citep{williams,bird2006transport}.
    \item The bulk viscosity vanishes if all components are monatomic, $\bar{C}_{a,v}=DR_U/2$.
    \item Introducing the thermal diffusivity $\alpha=\lambda/\rho C_p$ and the kinematic viscosity $\nu=\mu/\rho$, the Prandtl number becomes, 
\begin{align}
{\rm Pr} = \frac{\nu}{\alpha} = \frac{\omega_1(2-\omega)}{\omega(2-\omega_1)}.
\label{eqn:prandtl}
\end{align}
    


\item Using the equation of state of the mixture (\ref{eqn:eosIdealGas}) in (\ref{eq:mu}) and (\ref{eq:lambda}), relaxation parameters $\omega$ and $\omega_1$ are expressed in terms of  dynamic viscosity and thermal conductivity,  
\begin{align}
\omega &= \frac{2 P\delta t}{P\delta t+2 \mu},
\label{eqn:omega}\\
\omega_1 &= \frac{2 PC_p\delta t}{PC_p\delta t+2 \lambda}.
\label{eqn:omega1}
\end{align}
\end{itemize}

Finally, the dynamic viscosity $\mu$ and the thermal conductivity $\lambda$ of the mixture at any point is evaluated as a function of the local composition by using the methods described in \citet{wilke} and \citet{mathur}, respectively:

%
\begin{equation}
\mu = \sum_{a=1}^{M} \frac{\mu_a X_a}{\sum_{b=1}^{M} \phi_{ab} X_b},
\label{eqn:viscosityMixture}    
\end{equation}
where $\mu_a$ are the dynamic viscosity of the components while the dimensionless factor $\phi_{ab}$ is given by the equation,
\begin{equation}
\phi_{ab} = \frac{\left[1+\sqrt{\frac{\mu_a}{\mu_b} \sqrt{\frac{m_b}{m_a}}} \;\right]^2}{\sqrt{8} \sqrt{1+\frac{m_a}{m_b}}}.
\label{eqn:wilkePhi}    
\end{equation}

The thermal conductivity of the mixture $\lambda$ is calculated from the thermal conductivity of the components $\lambda_a$,
\begin{equation}
\lambda = \frac{1}{2}\left(\sum_{a=1}^{M} X_a \lambda_a + \frac{1}{\sum_{a=1}^{M} \frac{X_a}{\lambda_a}} \right).
\label{eqn:conductivityMixture}    
\end{equation}

\subsection{Realization on the standard lattice}
\label{sec:realization}

\subsubsection{Equilibrium and quasi-equilibrium}
In order to finalize the construction of the lattice Boltzmann equations for the mixture, one needs to specify the choice of the momentum and the energy lattices, and to provide the corresponding equilibrium and quasi-equilibrium populations. To that end, the single-component lattice Boltzmann models satisfying the moment constraints of Sec.\ \ref{sec:kinetic} are known in the literature. These employ higher-order lattices with a relatively large number of discrete velocities such as $D2Q49$ ($Q=49$ in two dimensions, \citet{frapolli2015}) or $D3Q39$ ($Q=39$ in three dimensions, \citet{frapolli2020}).

In this paper,  we develop the standard $D3Q27$ lattice realization  as in the above case of the species LBM of Sec.\ \ref{sec:27spec}. 
We thus consider a two-dimensional compressible single-component lattice Boltzmann model by \cite{hosseinCompressible} on the standard $D2Q9$ velocity set. 
Since the $D2Q9$ and $D3Q27$ belong to the same family of product-lattices, cf.\ \cite{karlin2010factorization}, it is natural to consider compressible LBM of \cite{hosseinCompressible} for our purpose.
Below, we extend the model of \cite{hosseinCompressible} to the three-dimensional $D3Q27$ discrete velocities set.

%


For the evaluation of the equilibrium  $g_i^{\rm eq}$ and of the quasi-equilibrium $g_i^*$ of the energy lattice, we proceed with the following ansatz $\mathcal{G}_i$, parameterized with a scalar $\theta\in[0,1]$, a scalar $M_0$, a vector $\bm{m}$ and a second-order tensor $\bm{M}$,
\begin{align}
&\mathcal{G}_i(\theta, M_0,\bm{m},\bm{M})= h_i(\theta, M_0,\bm{m},\bm{M}) + \bm{B}_{i}\cdot \bm{Z}(\theta,M_0,\bm{M}),
\label{eq:gansatz}\\
&h_i(\theta,M_0,\bm{m},\bm{M}) = w_i(\theta) \left(M_0 +  \frac{ \bm{m}\cdot \bm{c}_{i}}{\theta} + \frac{(\bm{M}-M_0 \theta \bm{I}): (\bm{c}_{i}\otimes\bm{c}_{i} - \theta \bm{I})}{2 \theta^2}\right).
\label{eqn:grad}
\end{align}
Here, the weights $w_i$ are calculated in the product form as,
\begin{equation}
w_i = w_{c_{ix}} w_{c_{iy}} w_{c_{iz}}
\label{eqn:weights}
\end{equation}
based on the fundamental triplet,
\begin{align}
w_{0} &= 1 - \theta, \ w_{1} = \frac{\theta}{2},\ w_{-1} = \frac{\theta}{2}.
\label{eqn:w0}
\end{align}
Furthermore in (\ref{eq:gansatz}), $\bm{Z}$ is a vector with the components
\begin{align}
\label{RcorrectionPrime}
Z_{\alpha} &= \frac{(1-3 \theta)}{2 \theta} (M_{\alpha \alpha} - \theta M_0).
\end{align}
Here, $M_{\alpha\alpha}$ is the diagonal component of the second-order tensor $\bm{M}$, 
while the components of vectors $\bm{B}_i$ are defined as follows:
\begin{equation} \label{eqn:Bi}
\begin{aligned}
B_{i \alpha} & = 1   &  \text{for} \;  c_i^2=0 \\
B_{i \alpha} & = -\frac{1}{2} \lvert c_{i \alpha} \rvert & \text{for} \; c_i^2=1  \\
B_{i \alpha} & = 0   &  \text{otherwise} .  
\end{aligned}
\end{equation}
Note that, when the parameter $\theta$ is set to the lattice reference temperature $\theta=1/3$, the term in (\ref{RcorrectionPrime}) vanishes and the remaining term (\ref{eqn:grad}) becomes the familiar second-order Grad's approximation \citep{grad1949}.
By construction, the form (\ref{eq:gansatz}) satisfies the moment relations, for any $\theta$:
\begin{equation}
    \sum_{i=0}^{Q-1}\left\{1, \bm{c}_i, \bm{c}_i\otimes\bm{c}_i\right\}\mathcal{G}_i(\theta, M_0,\bm{m},\bm{M})=\left\{M_0,\bm{m},\bm{M}\right\}.
\end{equation}

The equilibrium and the quasi-equilibrium populations $g_i^{\rm eq}$ and $g_i^{*}$ are defined with the help of the form (\ref{eq:gansatz}) by specifying the parameters $\theta=RT$, $M_0=\rho E$ and $\bm{M}=\bm{R}^{\rm eq}$ (\ref{eqn:geq2mom}) in both cases, and 
$\bm{m}=\bm{q}^{\rm eq}$ (\ref{eqn:geq1mom}) or $\bm{m}=\bm{q}^*$ (\ref{eqn:gstareq1mom}) for the equilibrium or the quasi-equilibrium, respectively, following \cite{hosseinCompressible}: 
\begin{align}
g_i^{\rm eq} &= \mathcal{G}_i( RT,\rho E,\bm{q}^{\rm eq},\bm{R}^{\rm eq}),
\label{eqn:gieq}
\\
g_i^{*} &= \mathcal{G}_i( RT,\rho E,\bm{q}^{*} , \bm{R}^{\rm eq}).
\label{eqn:gistar}
\end{align}
We shall now proceed with identifying the equilibrium of the momentum lattice and the modification of the lattice Boltzmann equation necessary for the $D3Q27$ model.

\subsubsection{Augmented lattice Boltzmann equation for the momentum lattice}
\label{sec:correction}
Equilibrium populations of the momentum lattice $f_i^{\rm eq}$ are evaluated in the conventional way with the help of the product-form (\ref{eq:prod}) and using $\bm{\xi}=\bm{u}$ and $\zeta=RT$, 
\begin{equation}
f_i^{\rm eq} = \rho \Psi_i(\bm{u},RT).
\label{eqn:feq}
\end{equation}
It is well known that the diagonal element of the equilibrium third order moment of the momentum lattice $Q_{\alpha \alpha \alpha}^{\rm eq}$ cannot satisfy the required moment relation (\ref{eqn:feq3mom}).
This happens due to the lattice constraint, $c_{i\alpha}^3=c_{i\alpha}$, which makes the diagonal third-order moments linearly dependent on the momentum, cf., e.\ g. \citep{karlin2010factorization}. Following  \citep{hosseinCompressible}, we consider the augmented lattice Boltzmann equation on the momentum lattice as follows,
\begin{align}
f_i(\bm{x}+\bm{c}_i,t+1) &= f_i(\bm{x},t) + \omega (f_i^{\rm eq} -f_i) + \bm{A}_{i}\cdot \bm{X},
\label{eqn:fwithcorrection}
\end{align}
%
%
where $\bm{X}$ is the vector with the components $\alpha=x,y,z$,
\begin{equation}
X_{\alpha} = -\partial_\alpha \left[ \left( \frac{1}{\omega}-\frac{1}{2} \right)\delta t
\partial_\alpha (\rho u_\alpha (1 - 3 R T) - \rho u_\alpha^3) \right],
\label{eqn:Xalpha}
\end{equation}
and where the components of vectors $\bm{A}_i$ are defined as,
\begin{equation} \label{eqn:Ai}
\begin{aligned}
A_{i \alpha} & = \frac{1}{2} c_{i \alpha}  &  \text{for} \;  c_i^2=1 \\ 
A_{i \alpha} & = 0   &  \text{otherwise}  
\end{aligned}
\end{equation}
This completes the realization of the mixture momentum and energy lattice Boltzmann equations on the standard $D3Q27$ lattice. In the next section we shall specify the coupling between the species and the mixture lattice Boltzmann subsystems.

\subsection{{Matching of mixture density and momentum: Weak and strong coupling}}
\label{sec:coupling}
\subsubsection{Weak coupling}
Summarizing, the lattice Boltzmann model for a compressible $M$-component mixture of ideal gas on the standard $D3Q27$ lattice consists of 
$M$ species lattices where the lattice Boltzmann equation is given by Eq.\ (\ref{eqn:SMlbm}), and the momentum and energy lattice Boltzmann equations (\ref{eqn:fwithcorrection}) and (\ref{eqn:g}). In total, the $M+2$ lattice Boltzmann equations are tightly coupled, as has been already specified above: The temperature from the energy lattice is provided to the species lattices through species equilibrium (\ref{eq:27eq}) and quasi-equilibrium (\ref{eq:27qeq}), but also in the Stefan--Maxwell temperature-dependent relaxation rates (\ref{eqn:smDiffusionCoefficients2}). On the other hand, the mass fractions from the species lattices are used to compute the mixture energy and enthalpy in the equilibrium and the quasi-equilibrium of the momentum and energy lattices. 
Another coupling is the input of species diffusion velocities into the quasi-equilibrium population of the energy lattice via the interdiffusion flux (\ref{eq:interdiffusion}). Finally, the momentum and the energy lattices are coupled in the standard way already present in the single-component setting. 
This entire set of interconnections between the species, and the momentum and energy lattices shall be termed the weak coupling.

\subsubsection{Strong coupling}

With the two subsystems, of the species and the mixture, first  constructed  independently from each other and after that being coupled in the way described above, we are left with two independent definitions of the mixture density and the mixture momentum: On the one hand,
the mixture density $\rho$ (\ref{eqn:f0momDensity}) and the mixture momentum $\rho\bm{u}$ (\ref{eqn:f1momMomentum}) are defined as the moments of the populations $f_i$.  
On the other hand, the same quantities are defined with the species populations as the sum of partial densities and partial momenta. 
The number of the conservation laws for the species subsystem is $M+D$, while for the mixture subsystem it is $D+2$. 
The total number of the conservation laws in the weakly coupled combined system is $M+2D+2$. Thus, the weakly coupled system is in excess of $D+1$ conservation laws as compared to the $M+D+1$ conservation laws of the mixture. 

We consider the mixture density and mixture momentum defined by the momentum lattice as primary. 
The over-determination is then resolved by replacing the density of a selected component (here, the $M^{th}$) $\rho_M$ by the deficit of density once the $M-1$ other components are taken into account. Similarly, the momentum of the $M^{th}$ component $\rho_M\bm{u}_M$ accounts for the deficit of the mixture momentum once the momenta of the other species are counted: 
\begin{align}
    \label{eq:rhoMATCH}
     \rho_M&=\rho-\sum_{a=1}^{M-1}\rho_{a}=\sum_{i=1}^{Q-1}f_i-\sum_{a=1}^{M-1}\sum_{i=1}^{Q-1}f_{ai},\\
      \rho_M\bm{u}_M&=\rho\bm{u}-\sum_{a=1}^{M-1}\rho_{a}\bm{u}_a=\sum_{i=1}^{Q-1}f_i\bm{c}_i-\sum_{a=1}^{M-1}\sum_{i=1}^{Q-1}f_{ai}\bm{c}_i.
    \label{eq:uMATCH}
\end{align}
In other words, the density and momentum of the $M^{th}$ component is no more an independent field but is slaved by the corresponding mixture quantities and the rest of the mixture composition. This means that the lattice Boltzmann equation for the $M^{th}$ component becomes purely relaxational, stripped of its conservation law. At the same time, the total momentum conservation is slaved by the momentum conservation of the momentum lattice. The number of independent conservation laws in this strongly coupled system is thus,
\begin{equation}
 \label{eq:DOF}
 (D+2)+ [(M+D)-1-D]=M+D+1,
\end{equation}
and corresponds to the locally conserved fields, $\rho_1,\dots,\rho_{M-1}$, $\rho$, $\rho\bm{u}$ and $\rho E$. While the assignment of the ``slaved" component $M$ is not unique, it is advisable to select the component which carries the majority of mass in the mixture.  

It should be emphasised that the strong coupling is only necessary to avoid an over-determined system. In practice, performing simulations under the weak coupling does not result in a deviation of mass or momentum except in cases where the boundary conditions introduce such a deviation. The matching of the density and the momentum of the mixture under the weak coupling condition thus highlights the intrinsic consistency of the proposed mean-field model.
\section{Results}
\label{sec:results}
\subsection{Overview of numerical implementation}
\label{sec:overviewnumerics}

In order to validate various physical aspects of the proposed lattice Boltzmann model, we consider four benchmarks as follows: 
\begin{itemize}
 \item Diffusion in a ternary gas mixture. This test case validates the Stefan--Maxwell diffusion and exhibits effects 
  such as a diffusion barrier and uphill diffusion, which cannot be captured by Fick's diffusion assumption.
  \item Diffusion in opposed jets. Here, we verify the coupling between the hydrodynamics and the diffusion model. 
  \item Speed of sound measurement in a mixture. This test case further validates the compressible model. 
  \item Three-dimensional Kelvin--Helmholtz instability. Finally, this canonical benchmark demonstrates the extension of the fully coupled model to three dimensions and therefore shows viability for complex flows including turbulence.
\end{itemize}
In all cases, we consider the temperature-independent specific heats for each component and set the reference temperature $T_0=0$ in (\ref{eq:specUa}), so that the internal energy is $U_a={C}_{a,v}T$ and the mixture internal energy is, $U=C_vT$, $C_v=\sum_{a=1}^MY_a{C}_{a,v}$, see Sec.\ \ref{sec:thermo2}.
The speed of sound $c_s$ is defined as,
\begin{equation}
c_s= \sqrt{\gamma R T},
\label{eqn:soundSpeed}
\end{equation}
where both the adiabatic exponent $\gamma=C_p/C_v$ and the specific gas constant $R$ depend on the mixture composition.
In what follows, we use the acoustic scaling: The speed of sound (\ref{eqn:soundSpeed}) at a specified reference composition (typically, at the equilibrium) and specified temperature is used to make velocity non-dimensional, unless otherwise stated. The characteristic 
length is given in the respective setup. Transport coefficients including dynamic viscosity, thermal conductivity and diffusivity were derived from the GRI-Mech 3.0 mechanism \citep{gri3}; acoustic scaling was used to convert these to lattice units.
Finally, the second-order accurate isotropic lattice operators proposed by \citet{thampi} were used for the evaluation of spatial derivatives in the correction to the heat flux (\ref{eq:corrFourier})
as well as in the isotropy correction  (\ref{eqn:Xalpha}).

\subsection{Diffusion in a ternary gas mixture}
\label{sec:ternary}

Classical experiment on diffusion in ternary mixture of hydrogen $H_2$, argon $Ar$ and methane $CH_4$ in a Loschmidt tube apparatus was performed by  \citet{arnold1967}; Results of the  experiment of \citet{arnold1967} were later analysed in depth by \citet{krishna}. Experiment of \citet{arnold1967} highlighted a number of, in part counter-intuitive, features of the Stefan--Maxwell diffusion. It is therefore natural to test our mixture model against the experiment of \citet{arnold1967}.


The strongly coupled version of the three-dimensional lattice Boltzmann model for Stefan--Maxwell diffusion was realized on a quasi-one-dimensional domain with 
$864 \times 1 \times 1$ grid points. In order to represent a closed tube, the bounce-back boundary condition was used for all populations at each end of the tube, while periodic boundary conditions were applied in the other two directions. The same initial composition of the mixture as in the experiment of \citet{arnold1967} was used; the setup was initialized  with a uniform atmospheric pressure and temperature $T=300\, K$.  It should be stressed that the temperature was not stipulated to be fixed during the simulation. Rather, the fully coupled thermo-hydrodynamic system maintained the isobaric and isothermal conditions by the initial and boundary conditions. As in the experiment, the evolution of the composition was represented by the average mole fraction of each species in the left and in the right halves of the tube. The non-dimensional time was used to represent the data, $t_{ND}=t/t_s$, where $t_s$ is the time required for the sound wave to traverse the domain in a reference equilibrium composition.

\begin{figure}
 \centering
 \includegraphics[keepaspectratio=true,width=0.9
 \textwidth]{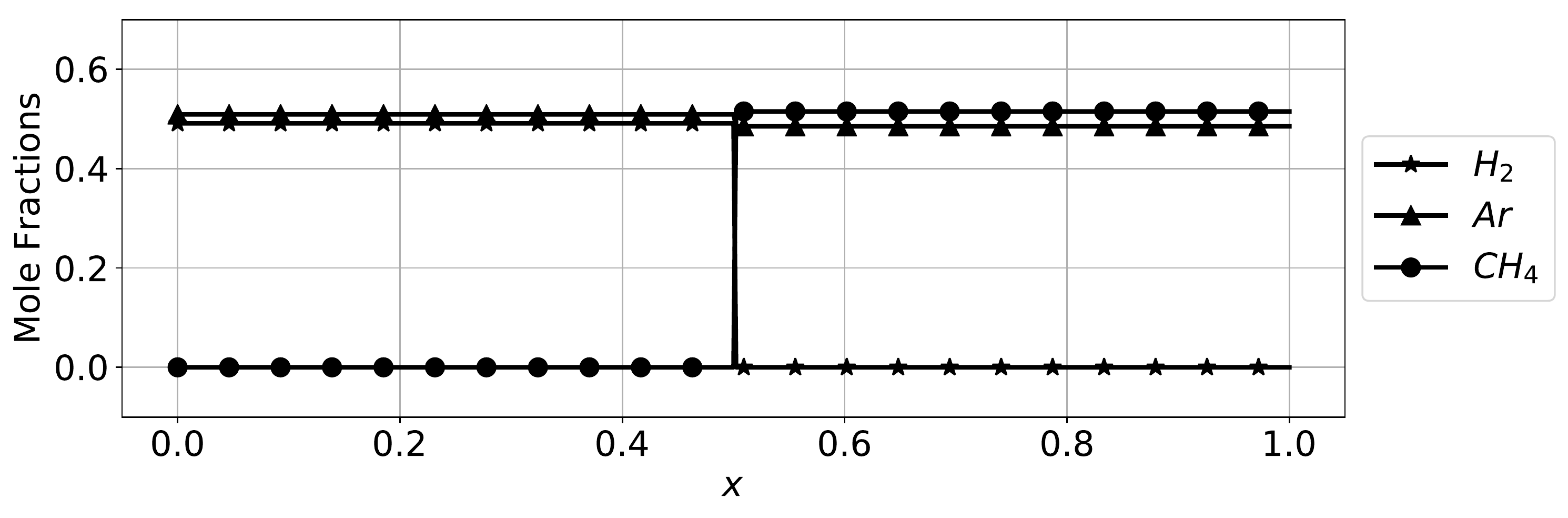}
 \caption{Diffusion in ternary mixture, case $1T$ \citep{arnold1967}. Mole fractions of hydrogen $H_2$, argon $Ar$ and methane $CH_4$ along the length of the tube as described by the initial conditions of Eq.\ (\ref{ternaryConditions}).}
 \label{fig:x1InitialState}
\end{figure}

In our first numerical experiment,  the mole fractions of the species in the left and in the right halfs of the tube were initiated as in the case $1T$  of \citet{arnold1967}; see Fig.\ \ref{fig:x1InitialState}:
\begin{equation}
\begin{aligned}
& {\rm Left}  & X_{H_2}=0.491, X_{Ar}=0.509, X_{CH_4}=0.000 \\
& {\rm Right} & X_{H_2}=0.000, X_{Ar}=0.485, X_{CH_4}=0.515
\label{ternaryConditions}
\end{aligned}
\end{equation}

\begin{figure}
 \centering
 \includegraphics[keepaspectratio=true,width=0.9
 \textwidth]{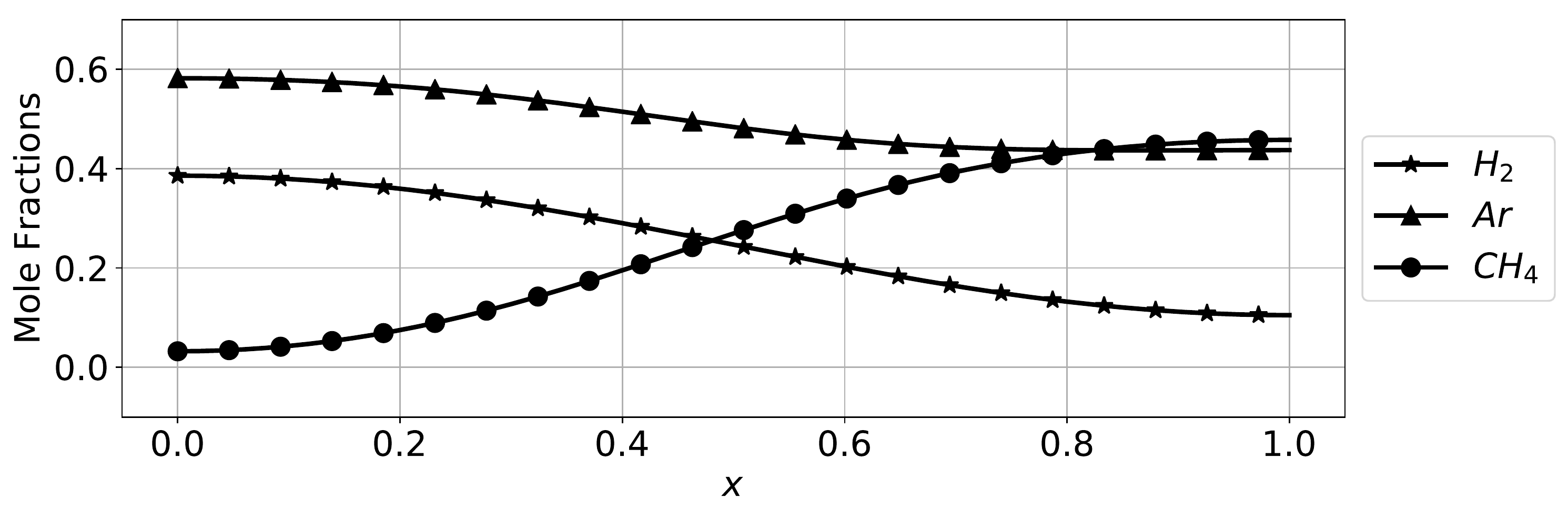}
 \caption{Case $1T$. Mole fractions of hydrogen $H_2$, argon $Ar$ and methane $CH_4$ along the length of the tube during uphill diffusion of $Ar$ at $t_{ND}=179.52$.}
 \label{fig:x1TduringUphillDiffusion}
\end{figure}
\begin{figure}
 \centering
 \includegraphics[keepaspectratio=true,width=0.9
 \textwidth]{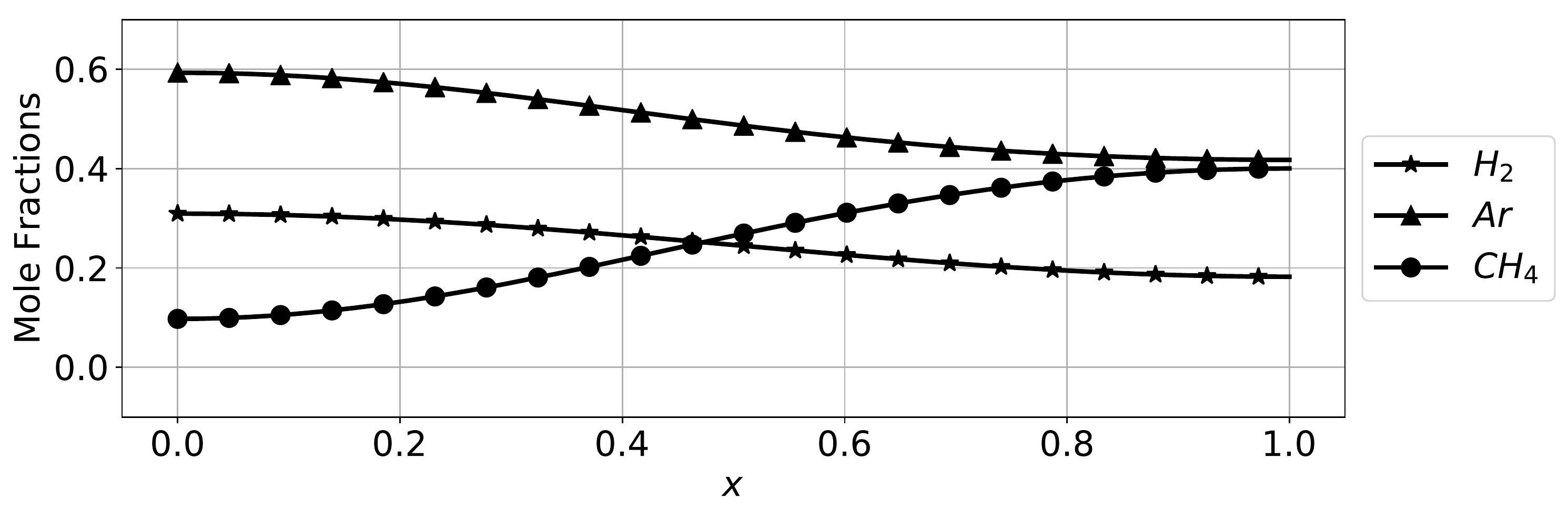}
 \caption{Case $1T$. Mole fractions of hydrogen $H_2$, argon $Ar$ and methane $CH_4$ along the length of the tube at the diffusion barrier, $t_{ND}=378.98$.}
 \label{fig:x1TatDiffusionBarrier}
\end{figure}
\begin{figure}
 \centering
 \includegraphics[keepaspectratio=true,width=0.9
 \textwidth]{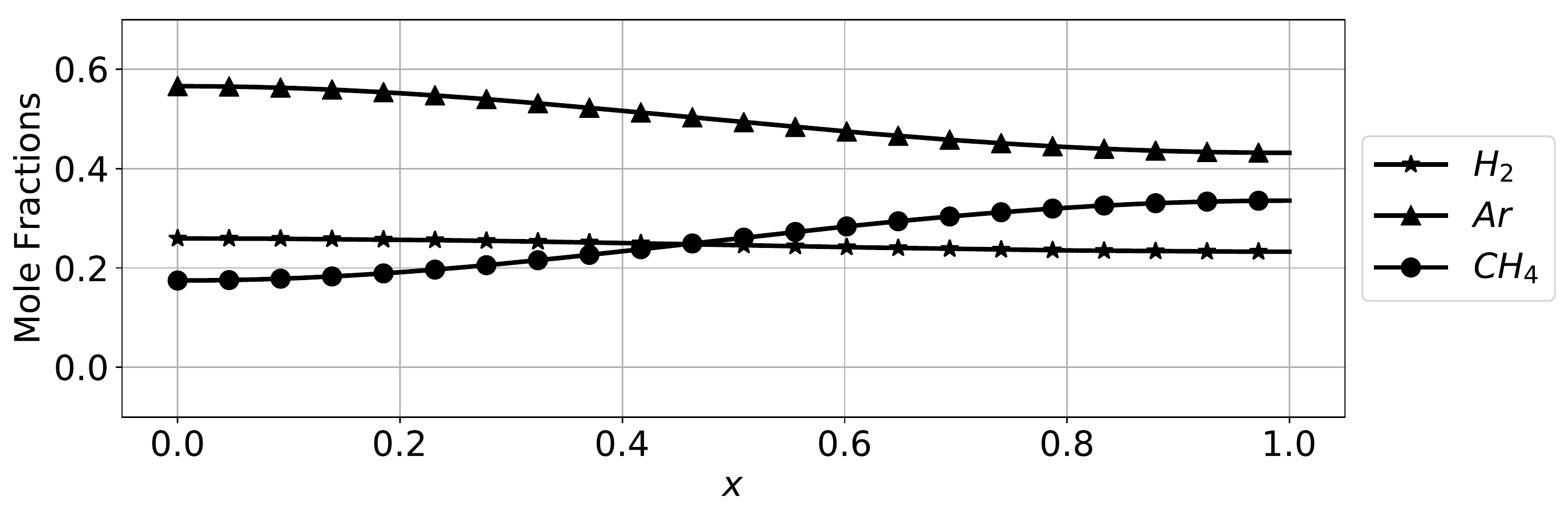}
 \caption{Case $1T$. Mole fractions of hydrogen $H_2$, argon $Ar$ and methane $CH_4$ along the length of the tube during Fickian diffusion of argon after the diffusion barrier, $t_{ND}=777.91$.}
 \label{fig:x1TduringFickianDiffusion}
\end{figure}

\begin{figure}
 \centering
 \includegraphics[keepaspectratio=true,width=0.9
 \textwidth]{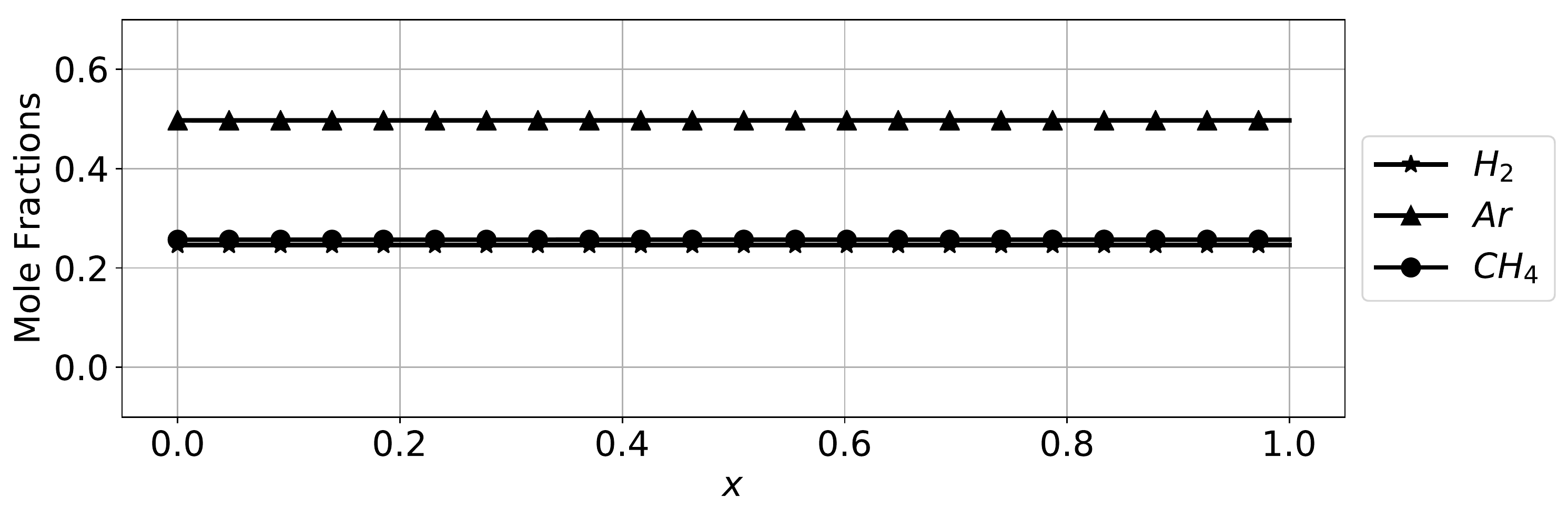}
 \caption{Case $1T$. Mole fractions of hydrogen $H_2$, argon $Ar$ and methane $CH_4$ along the length of the tube at the the steady state, $t_{ND}=6323.09$.}
 \label{fig:x1TatSteadyState}
\end{figure}

\begin{figure}
 \centering
 \includegraphics[keepaspectratio=true,width=0.7\textwidth]{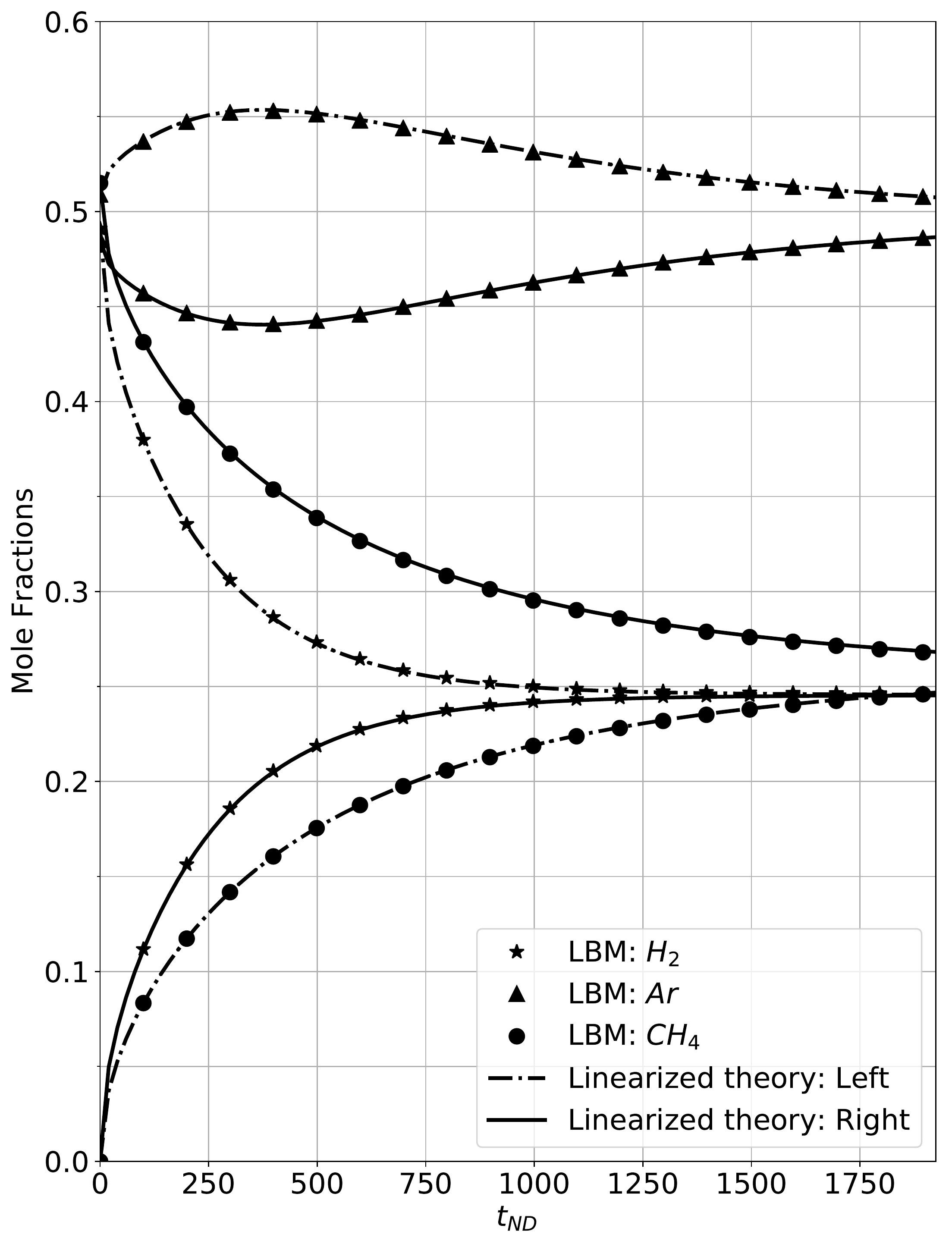}
 \caption{Diffusion in ternary mixture, case $1T$ \citep{arnold1967}, Eq.\ (\ref{ternaryConditions}). Averaged mole fractions of hydrogen $H_2$, argon $Ar$ and methane $CH_4$ in the left half and the right half sections of the tube. The figure shows reverse diffusion of argon. Symbol: present simulation; Line: theory \citep{arnold1967}.}
 \label{fig:moleFractionsAr}
\end{figure}

%

%
%
Time evolution of hydrogen $H_2$ and of methane $CH_4$ follows  Fick's diffusion law: the species from the higher concentration side reduce in mole fractions as they move towards the low concentration side. Thus, $CH_4$ can be seen moving from right to left and $H_2$ in the opposite direction, both species eventually attaining a uniform concentration. 

However, the behaviour of argon $Ar$ cannot be explained by Fick's law. Although $Ar$ has a negligible concentration gradient due to the initial conditions (\ref{ternaryConditions}), it does start diffusing, see Fig.\ \ref{fig:x1TduringUphillDiffusion}. This phenomenon was termed \emph{osmotic diffusion} by \cite{toor1957}. Osmotic diffusion is said to occur when the rate of diffusion of a component is not zero even though its concentration gradient is negligible;  this would correspond to an \emph{infinite} Fick's diffusivity.
The concentration of $Ar$ keeps on growing in the left section even though its concentration is higher in the left section itself, see Fig.\ \ref{fig:x1TduringUphillDiffusion}.  The effect was termed  \emph{uphill diffusion} (or reverse diffusion) in \cite{toor1957} because the component diffuses in the direction of increase of its concentration; in Fick's picture this would amount to \emph{negative} diffusivity.
The reverse diffusion is seen to proceed for some time and then flattens at $t_{ND} \approx 400$. At this point in time, an appreciable concentration gradient is built up but the diffusion is negligible, see Fig.\  \ref{fig:x1TatDiffusionBarrier}. The effect was termed a  \emph{diffusion barrier} in  \citep{krishna}, the point at which the diffusion rate of a component vanishes even though its concentration gradient does not. This would mean \emph{zero} Fick's diffusivity. After the diffusion barrier, the ordinary Fick's diffusion sets in and proceeds downhill of  the concentration gradient until the uniform steady state is reached, see Figs.\ \ref{fig:x1TduringFickianDiffusion} and \ref{fig:x1TatSteadyState}.    

Fig.\ \ref{fig:moleFractionsAr} shows the evolution of the average  mole fractions of the species in the left and in the right halves of the tube. 
The effects just mentioned were observed in the experiment of \citet{arnold1967} and in an earlier  similar experiment by \citet{duncan1962}.
\citet{krishna} provided explanation by drawing an analogy between the frictional drag and binary diffusion coefficients of pairs of species. According to \citet{krishna}, the Stefan-Maxwell diffusivity plays a role of an inverse drag coefficient. 
The binary diffusion coefficient between $Ar$ and $H_2$ is $8.14543 \times 10^{-5}  m^2/s$ while that between $Ar$ and $CH_4$ is $2.17321 \times 10^{-5} m^2/s$. This means that the frictional drag exerted on $Ar$ by $CH_4$ is much greater than that exerted on $Ar$ by $H_2$. Thus, $CH_4$ drags $Ar$ along with it during the initial period when the flux of $CH_4$ from right to left is large, causing the uphill diffusion of $Ar$. The transport of $CH_4$ from right to left eventually reduces because the driving force causing it reduces due to the reduction in concentration gradient of $CH_4$. At the same time, the increasing concentration of $Ar$ creates a driving force for $Ar$ to diffuse  downhill.  A balance is reached at the point of diffusion barrier after which the drag force caused by $CH_4$ is overcome and $Ar$ starts diffusing downhill of its concentration in a  Fick's fashion.   

It is apparent from Fig.\ \ref{fig:moleFractionsAr} that the lattice Boltzmann simulation  was able to correctly capture the experimentally observed phenomena. For a more quantitative assessment, we compare simulation results with 
the linearized theory of multicomponent mass transfer proposed in \citet{arnold1967}. The theory relies upon a semi-analytical solution of the one-dimensional diffusion equations for the average mole fractions using linearized Stefan--Maxwell relation for the diffusion fluxes, and was shown to match the experiment in a quantitative fashion \citep{arnold1967}. For the purpose of this study, we numerically solved the equations of the linearized theory using Python.
As is evident from Fig.\ \ref{fig:moleFractionsAr}, the results of the simulation agree well with the linearized theory, both in terms of magnitudes of the mole fractions as well as the time.

\begin{figure}
 \centering
 \includegraphics[keepaspectratio=true,width=0.7\textwidth]{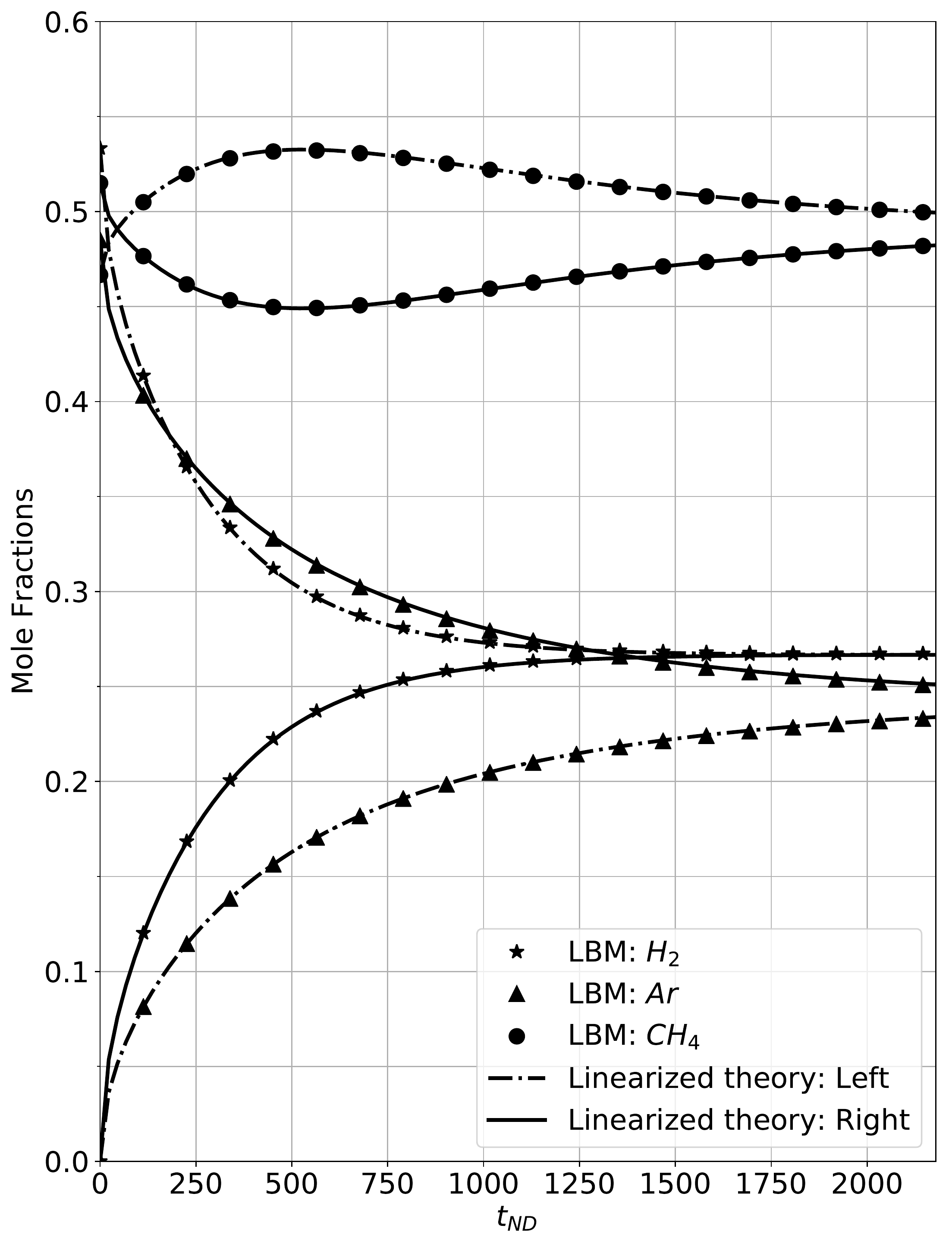}
 \caption{Diffusion in ternary mixture, case $2T$ \citep{arnold1967}, Eq.\ (\ref{eqn:ternaryConditionsCH4}). Averaged mole fractions of hydrogen $H_2$, argon $Ar$ and methane $CH_4$ in the left half and the right half sections of the tube. The figure shows reverse diffusion of methane. Symbol: present simulation; Line: theory \citep{arnold1967}.}
 \label{fig:moleFractionsCH4}
\end{figure}

\begin{figure}
 \centering
 \includegraphics[keepaspectratio=true,width=0.7\textwidth]{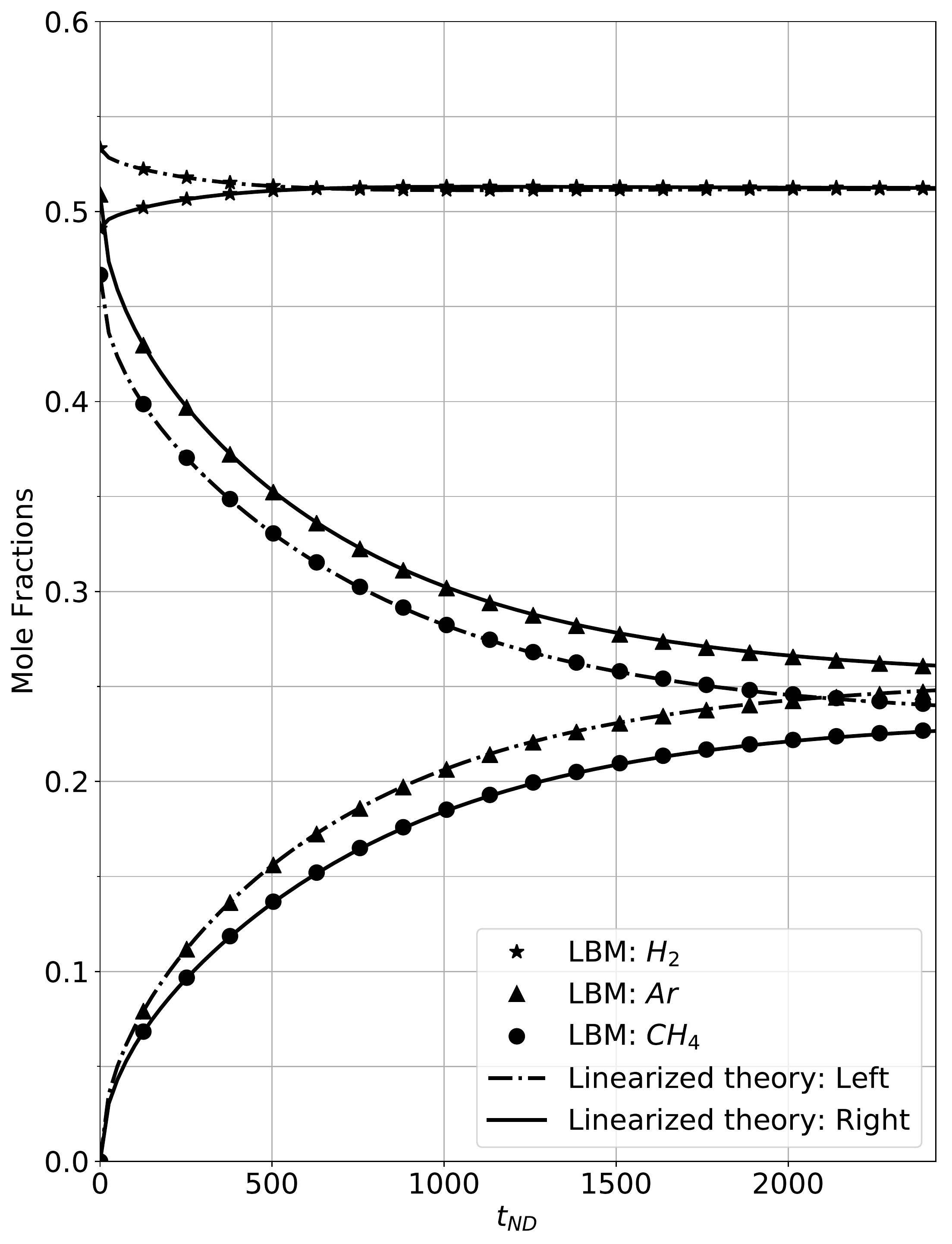}
 \caption{Diffusion in ternary mixture, case $3T$ \citep{arnold1967}, Eq.\ (\ref{eqn:ternaryConditionsH2}). Averaged mole fractions of hydrogen $H_2$, argon $Ar$ and methane $CH_4$ in the left half and the right half sections of the tube.. The figure shows near-Fickian diffusion of hydrogen. Symbol: present simulation; Line: theory \citep{arnold1967}.}
 \label{fig:moleFractionsH2}
\end{figure}

\begin{figure}
 \centering
 \includegraphics[keepaspectratio=true,width=0.7\textwidth]{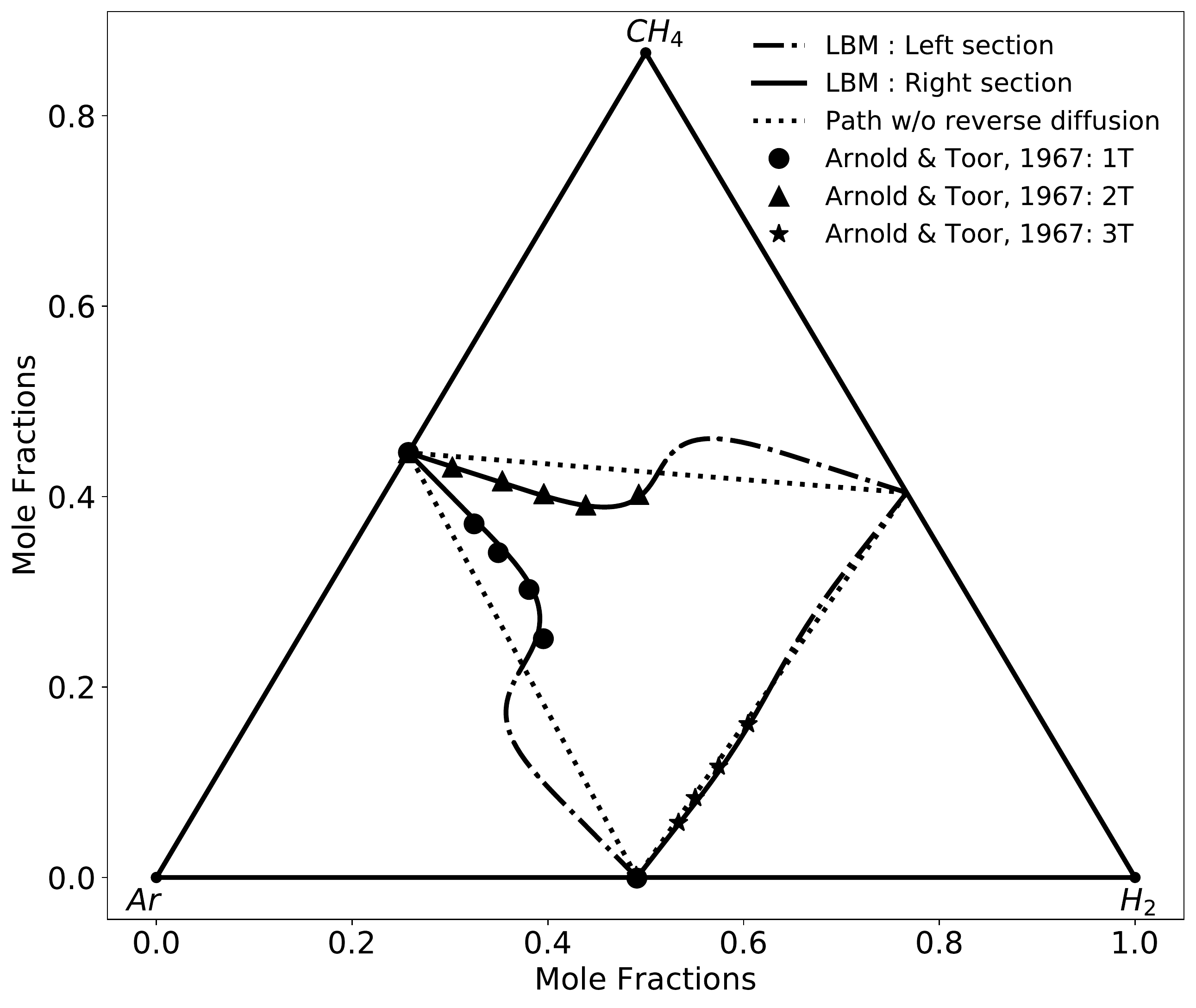}
 \caption{Composition path of the averaged mole fractions of $H_2$, $Ar$ and $CH_4$ in the left half and the right half section of the tube. The composition path shows three different cases, each with a reverse diffusion of $Ar$, $CH_4$ and a nearly Fickian diffusion of hydrogen. Lines: present simulation; Symbol: experiment of \cite{arnold1967}.}
 \label{fig:compositionPath}
\end{figure}

Continuing along the lines of the experimental study, the simulation was repeated with a different set of initial conditions, corresponding to the case $2T$ of \cite{arnold1967}:
\begin{equation}
\begin{aligned}
& {\rm Left}  & X_{H_2}=0.512, X_{Ar}=0.000, X_{CH_4}=0.448 \\
& {\rm Right} & X_{H_2}=0.000, X_{Ar}=0.485, X_{CH_4}=0.515
\label{eqn:ternaryConditionsCH4}
\end{aligned}
\end{equation}
According to the  theory of the inverse relation between mass diffusivity and drag \citep{krishna}, methane  should now exhibit uphill diffusion due to the flux of argon. This is indeed what was seen in the experiments of \cite{arnold1967} as well as in our simulations, see Fig.\ \ref{fig:moleFractionsCH4}. Simulations are in good agreement with the linearized theory.

A final but equally important situation is the one marked as case $3T$ in the experiment  \citep{arnold1967}, 
%
\begin{equation}
\begin{aligned}
& {\rm Left}  & X_{H_2}=0.512, X_{Ar}=0.000, X_{CH_4}=0.448 \\
& {\rm Right} & X_{H_2}=0.491, X_{Ar}=0.509, X_{CH_4}=0.000
\label{eqn:ternaryConditionsH2}
\end{aligned}
\end{equation}
The binary  diffusivity between $Ar$ and $H_2$ is $8.14543 \times 10^{-5}  m^2/s$ while that between $CH_4$ and $H_2$ is $7.37433 \times 10^{-5} m^2/s$. The diffusivities are comparable and thus the interaction of $H_2$ with $Ar$ is very similar to the interaction of $H_2$ with $CH_4$. In Fig.\ \ref{fig:moleFractionsH2}, the results from the lattice Boltzmann simulation as well as the linearized theory show a nearly-Fickian diffusion of $H_2$. Hydrogen however does show a small but nevertheless clear tendency to accumulate in the right half of the tube. This is possibly due to a slightly greater drag exerted on $H_2$ by  $CH_4$  thanks to a somewhat smaller diffusivity between the pair.

For an additional validation, we compare the composition map of the simulation with the composition map of the experiment, since the composition map for the Stefan--Maxwell diffusion is independent  of time \citep{duncan1962}. Fig.\ \ref{fig:compositionPath} verifies that the composition paths of the simulations agree well with that of all the three experiments of \citet{arnold1967}. The composition path that would be followed by a purely Fickian diffusion is also marked in Fig.\ \ref{fig:compositionPath} as `Path w/o reverse diffusion' for the purpose of contrast. The setups eventually attain a homogeneous composition at the `Equilibrium' points on the composition map located  midway of the Fickian lines, at the intersection with the Stefan--Maxwell trajectory. It can be again seen in the composition map that even for the case $3T$, the diffusion path of  hydrogen is almost yet  not purely Fickian.


\subsection{Diffusion in opposed jets}
\label{sec:OpposedJets}

\begin{figure}
 \centering
 \includegraphics[keepaspectratio=true,width=0.8\textwidth]{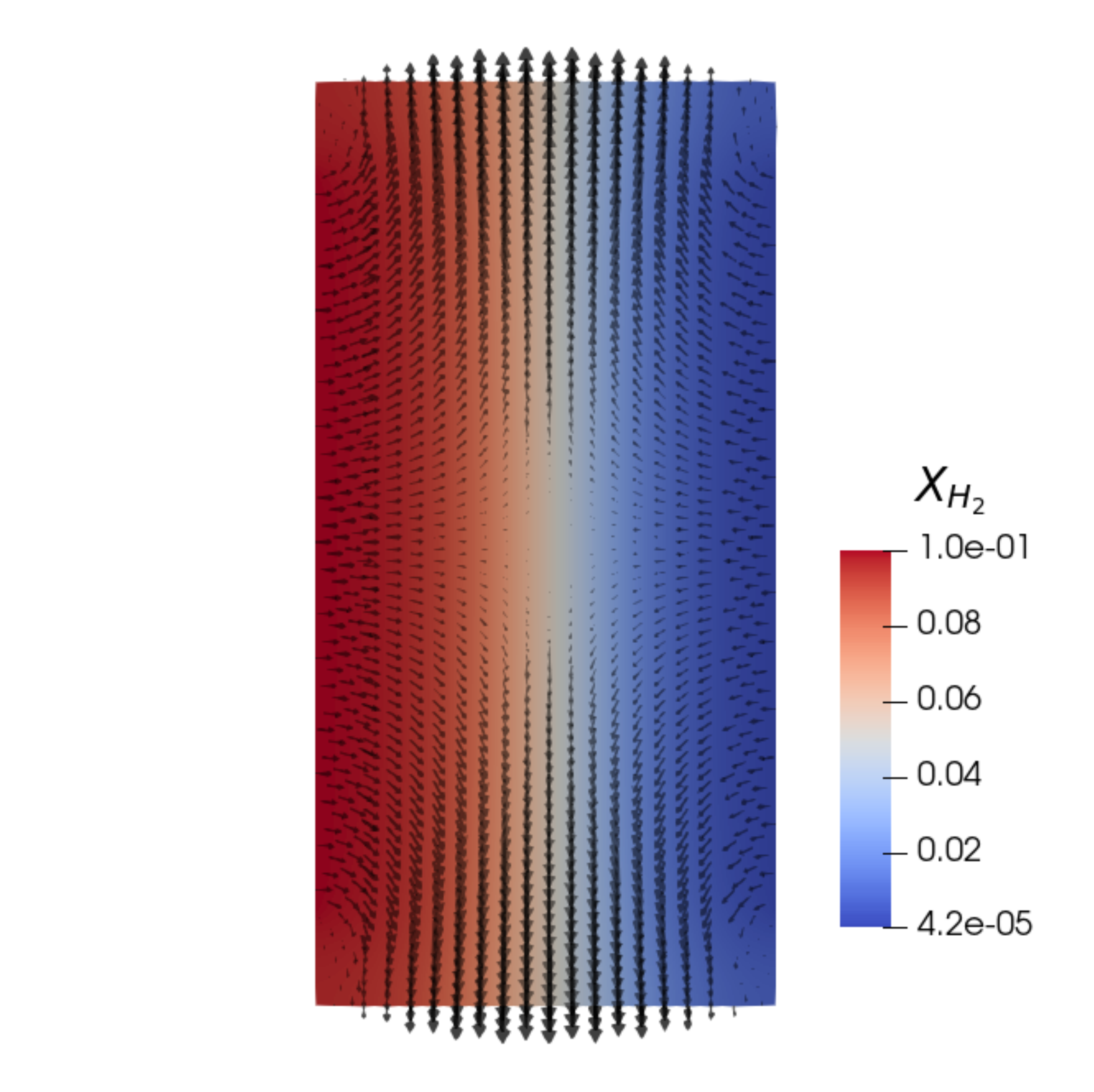}
 \caption{Contour of the mole fraction of $H_2$ and vectors of velocity at steady state for the opposed jets setup. The velocity vectors are scaled by the magnitude of the velocity.}
 \label{fig:ojSetup}
\end{figure}

In order to assess the coupling between the diffusion and the hydrodynamic systems we consider the case of planar opposed jets. The setup and boundary conditions are similar to that studied in \citet{salvoMulti}. It consists of two facing jets of fluid with equal momentum and different compositions. As shown in Fig.\ \ref{fig:ojSetup}, the simulation is performed on a grid of size $L_x \times L_y \times L_z = 200 \times 400 \times 1$ points, with the distance between the nozzles $L_x=200$. For the inlets, the incoming populations are replaced by the equilibrium distributions while the the outlets are modelled by making the derivative normal to the boundary zero. For the solid vertical boundaries at $y < 0.1 L_y$ and $y > 0.9 L_y$, a free-slip boundary condition is used.
The compositions of the jet streams are,
\begin{equation}
\begin{aligned}
& {\rm Left}  & X_{H_2}=0.1, X_{N_2}=0.85, X_{O_2}=0.0, X_{H_2O}=0.05 \\
& {\rm Right}  & X_{H_2}=0.0, X_{N_2}=0.90, X_{O_2}=0.1, X_{H_2O}=0.00
\label{eqn:ojInitialConditions}
\end{aligned}
\end{equation}

\begin{figure}
 \centering
 \includegraphics[keepaspectratio=true,width=0.8\textwidth]{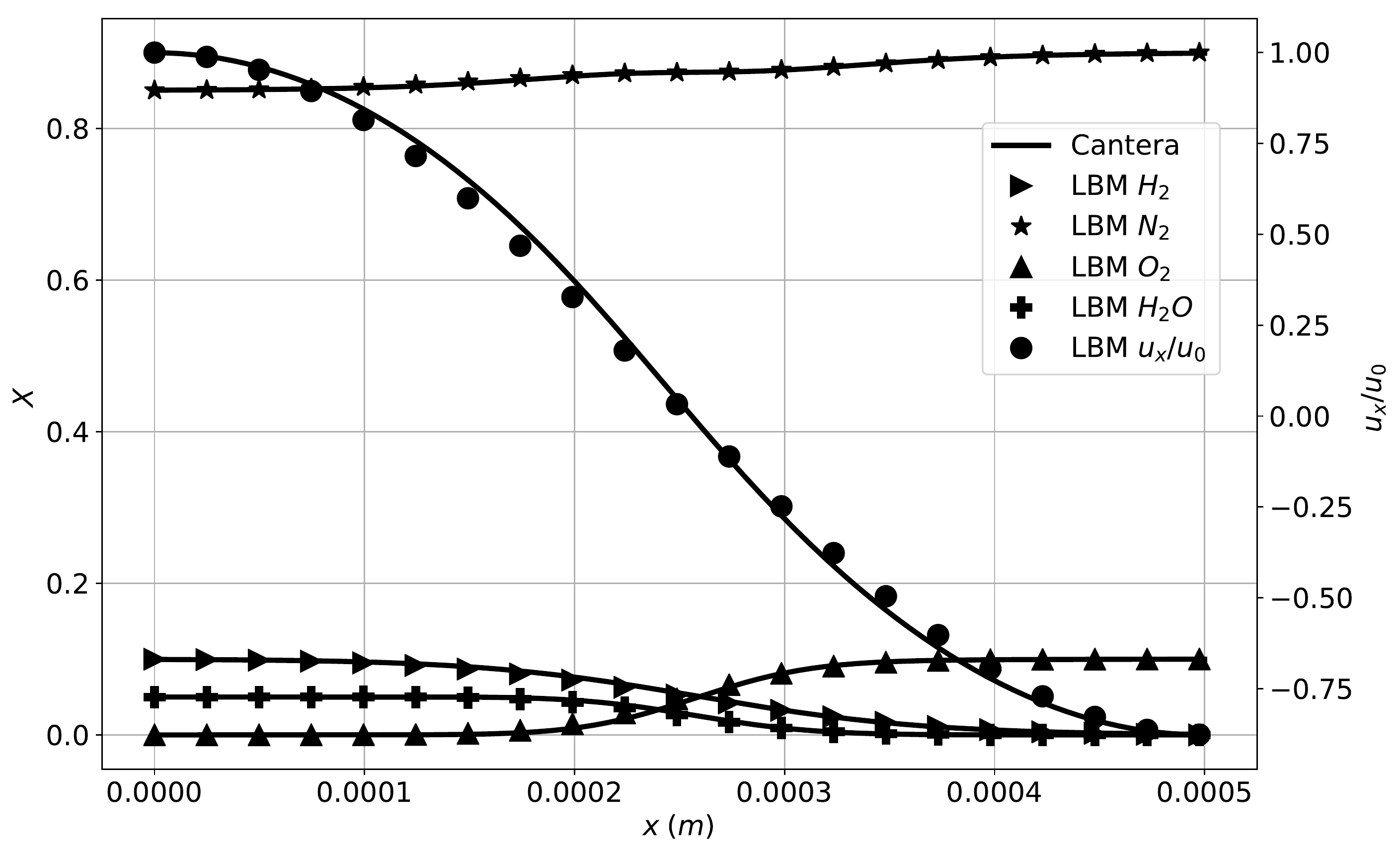}
 \caption{Mole fractions of $H_2$, $N_2$, $O_2$ and $H_2O$ and flow velocity at stagnation line.}
 \label{fig:ojResults}
\end{figure}

The solution of the present model at steady state is compared to the solution produced by the `CounterflowDiffusionFlame' function of the open source package Cantera \citep{cantera}. This function computes a steady state solution to counter-flow diffusion flame using a reduced one-dimensional similarity solution, as derived in section 6.2 of \citet{Kee}. 
In order to get a solution comparable with the Stefan--Maxwell formulation of the lattice Boltzmann model, the reactions are turned off in Cantera and the transport model is set to `Multi', which accounts for the pairwise diffusion between the species.
As can be seen in Fig.\ \ref{fig:ojResults}, the LBM solution for the mole fractions of all the components as well as the scaled velocity agree well with solution produced by Cantera. 
It should be noted that this test case is regarded severe in \citet{salvoMulti} since the diffusion of the components proceeds against the velocity of the bulk flow. For example, hydrogen and water from the left nozzle diffuse against the bulk flow on the right side, upstream towards the right nozzle. The good agreement of the results indicates that the coupling of the Navier--Stokes and the Stefan--Maxwell models is consistent and correct.

\subsection{Speed of sound}
\label{sec:soundSpeedTest}

\begin{figure}
 \centering
 \includegraphics[keepaspectratio=true,width=0.8\textwidth]{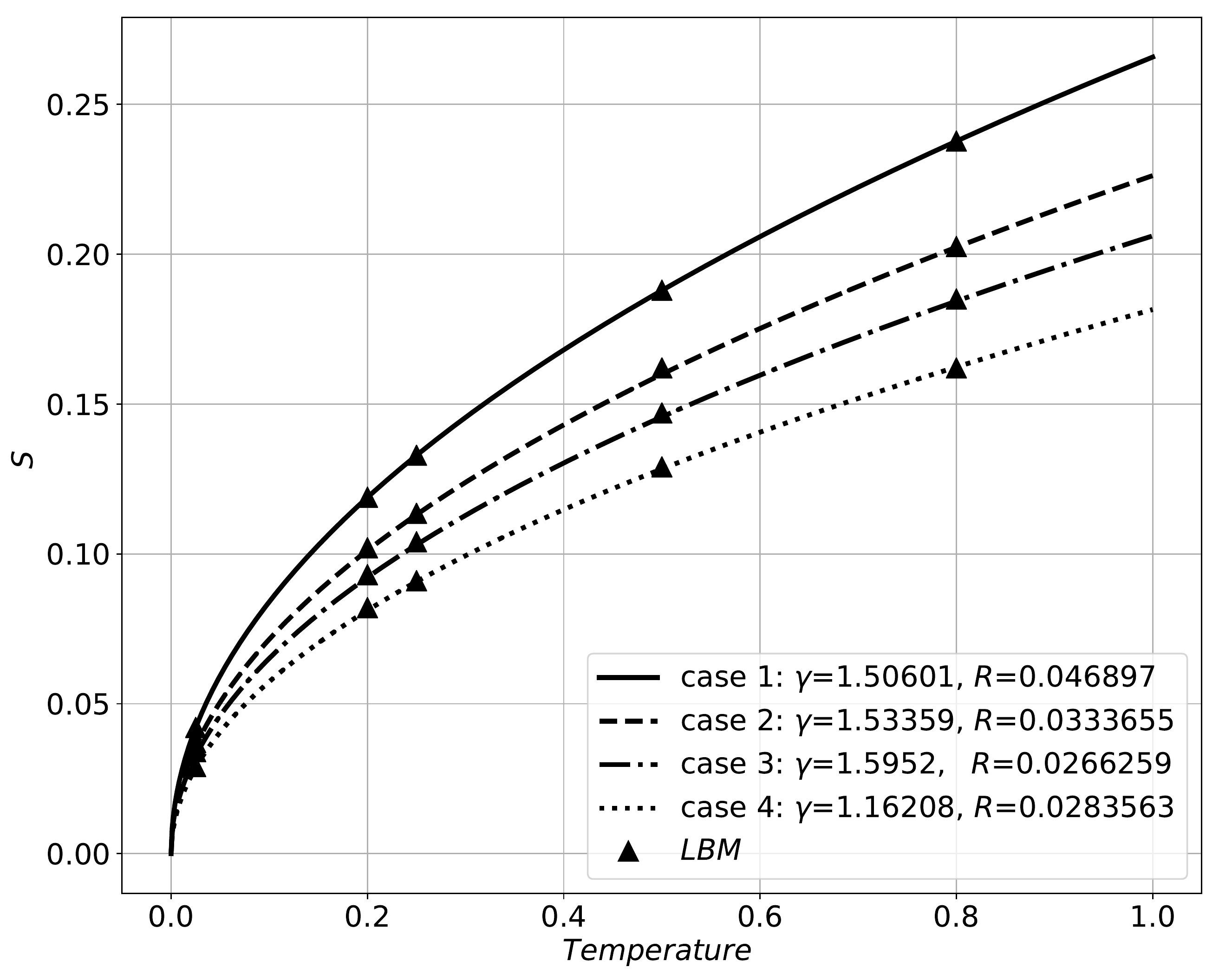}
 \caption{Speed of sound for different compositions (\ref{eqn:soundTestConditions}). Symbol: Simulation; Line: Theory, Eq.\ (\ref{eqn:soundSpeed}).}
 \label{fig:soundSpeed}
\end{figure}

As a standard test for a compressible flow LBM, we verify that the model correctly reproduces the speed of sound $c_s$ (\ref{eqn:soundSpeed}). 
The speed of sound was measured  for the following four compositions $S1$--$S4$:
\begin{equation}
\begin{aligned}
(S1) \;& {\rm R=0.046897} & X_{H_2}=0.491, X_{Ar}=0.509, X_{CH_4}=0.000 \\
(S2) \;& {\rm R=0.0333655} & X_{H_2}=0.200, X_{Ar}=0.700, X_{CH_4}=0.100 \\
(S3) \;& {\rm R=0.026625} & X_{H_2}=0.000, X_{Ar}=0.900, X_{CH_4}=0.100 \\
(S4) \;& {\rm R=0.0283563} & X_{H_2}=0.200, X_{Ar}=0.100, X_{C_3 H_8}=0.700
\label{eqn:soundTestConditions} 
\end{aligned}
\end{equation}
Composition S1 in (\ref{eqn:soundTestConditions}) is chosen from one of the cases in section (\ref{sec:ternary}) whereas the case S3 is chosen arbitrarily in order to test for the sound speed in a composition with a considerable difference in mole fractions. The cases S2 and S4  are chosen to verify the speed of sound in a ternary mixture and in the presence of heavier gases such as propane, respectively. The test is performed by tracking a small perturbation in pressure $\Delta P=10^{-5}$ at a specified temperature $T$. Fig.\ \ref{fig:soundSpeed} compares the measured speed of the propagation of the perturbation  with the theoretical speed of sound prediction  (\ref{eqn:soundSpeed}). The lattice Boltzmann model correctly recovered the sound speed over a tested  range of temperatures from $T_{\rm min}=0.025$ to $T_{\rm max}=0.8$, in lattice units.
The tested temperature range is characterized by the ratio of the temperatures, $T_{\rm max}/T_{\rm min}=32$ which is sufficiently large for many applications.
Temperature  between $T=0.2$ to $T=0.5$ in lattice units was used for most of the simulations presented in this paper.   

\subsection{Kelvin--Helmholtz instability}
\label{sec:khi}


Without a pretence of an in-depth study of shear layers instabilities in this paper, the final example presents a three-dimensional simulation of the classical Kelvin--Helmholtz instability (KHI) in order to validate the proposed model towards its possible use for  high Reynolds number simulations.
Similar to the setup in \citet{sanKHI}, we simulate the Kelvin-Helmholtz instability in a periodic domain of the size $L_x \times L_y \times L_z = 800 \times 800 \times 200$ lattice grid points. 
The domain is split into three sections in the flow-normal direction, where the initial conditions for a two-component mixture of nitrogen $N_2$ and water vapor $H_2O$ are as follows:
%
%
\begin{equation}
\begin{aligned}
& u_x=0.1 M_c, & X_{H_2O}=0.9,\quad& X_{N_2}=0.1,   \quad \text{for } 0 \leq y < 0.25 L_y,\\
& u_x=-0.1M_c, & X_{H_2O}=0.1,\quad& X_{N_2}=0.9,   \quad \text{for } 0.25L_y \leq y < 0.75 L_y,\\
& u_x=0.1M_c, & X_{H_2O}=0.9, \quad& X_{N_2}=0.1,   \quad \text{for } 0.75L_y \leq y \leq L_y.
\label{eqn:icKHI} 
\end{aligned}
\end{equation}
The velocity in the normal direction $u_y$ and in the span-wise direction $u_z$ is given by, respectively,
\begin{align}
u_y&=2 \lvert u_x \rvert 0.01 \sin(2 \pi x/L_x), 
\label{eqn:uyKHI}\\
u_z&=2 \lvert u_x \rvert 0.01 \sin(2 \pi z/L_z).
\label{eqn:uzKHI}
\end{align}
The initial composition of the binary mixture (\ref{eqn:icKHI}) is so chosen as to equilibrate at the $50/50$ equilibrium composition 
$X_{N_2}=X_{H_2O}=0.5$ in the absence of any flow.
The initial condition (\ref{eqn:uyKHI}) and (\ref{eqn:uzKHI}) further introduces a small perturbation in both the normal and the span-wise directions with a wavelength equal to the length of the domain and a magnitude of one percent of the relative shear velocity. As defined by \citet{leepKHI}, the convective Mach number $M_c$ is the Mach number relative to the frame of reference of the simulation. The relative Mach number $M_r$ based on the relative velocity across the shear layers is $M_r=0.2$ according to the initial conditions (\ref{eqn:icKHI}). The Reynolds number with respect to the viscosity of the bottom-most layer, with $M_r$ as the velocity scale and $L_y$ as the length scale is $Re=11963.46$. The mole fractions are chosen $0.1$ and $0.9$  to make the test more severe. 
We define an eddy turnover time $t_e=L_x/U_r$, with  the initial relative velocity $U_r=2 \lvert u_x \rvert$.
\begin{figure}
 \centering
 \includegraphics[keepaspectratio=true,width=0.9\textwidth]{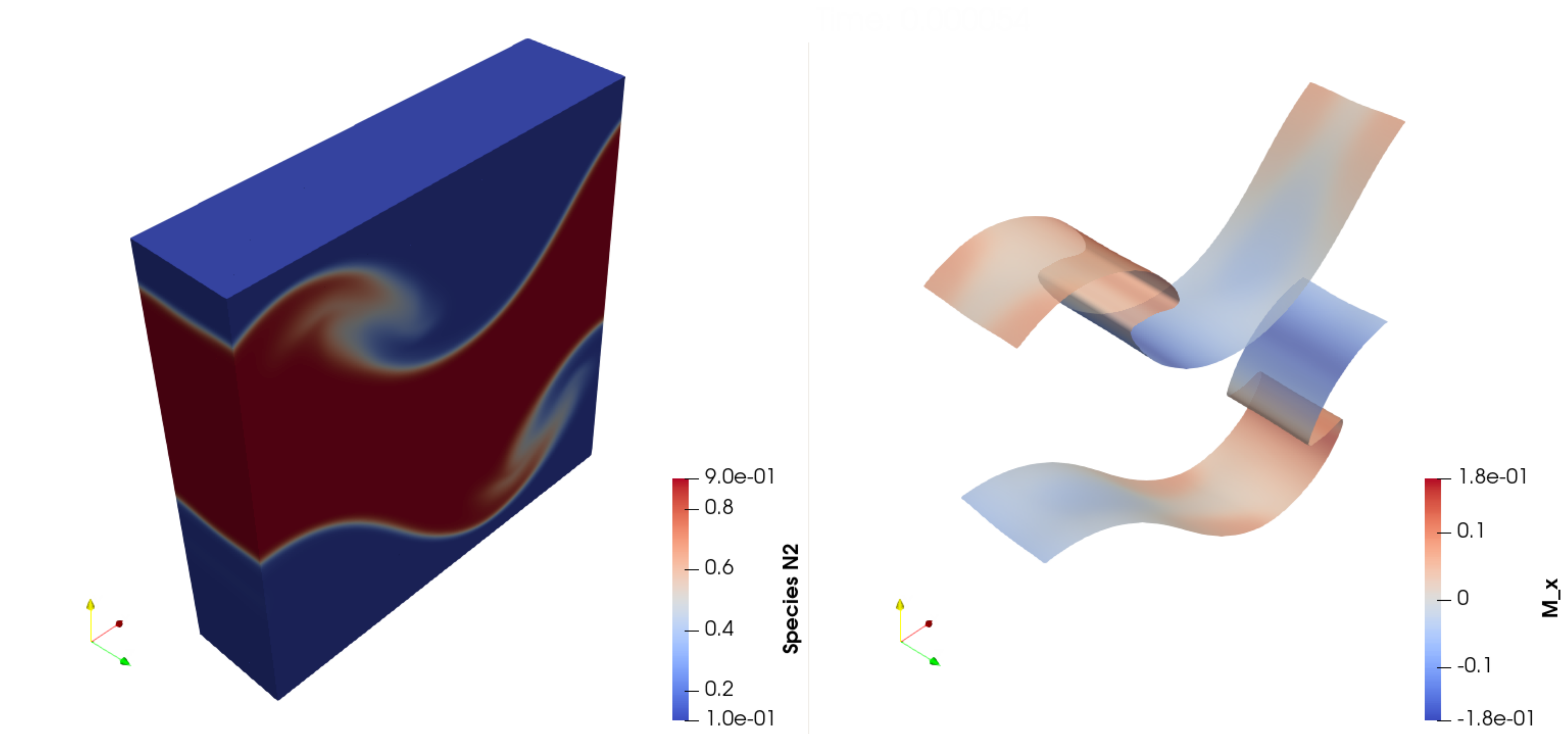}
 \caption{Contour of mole fraction of nitrogen $N_2$ (left) and iso-surface of the equilibrium concentration of nitrogen $X_{N_2}=0.5$ colored with the Mach number in the $x$-direction  (right) at time $t_e=2.2521$.}
 \label{fig:frame17_N2X}
\end{figure}
\begin{figure}
 \centering
 \includegraphics[keepaspectratio=true,width=0.9\textwidth]{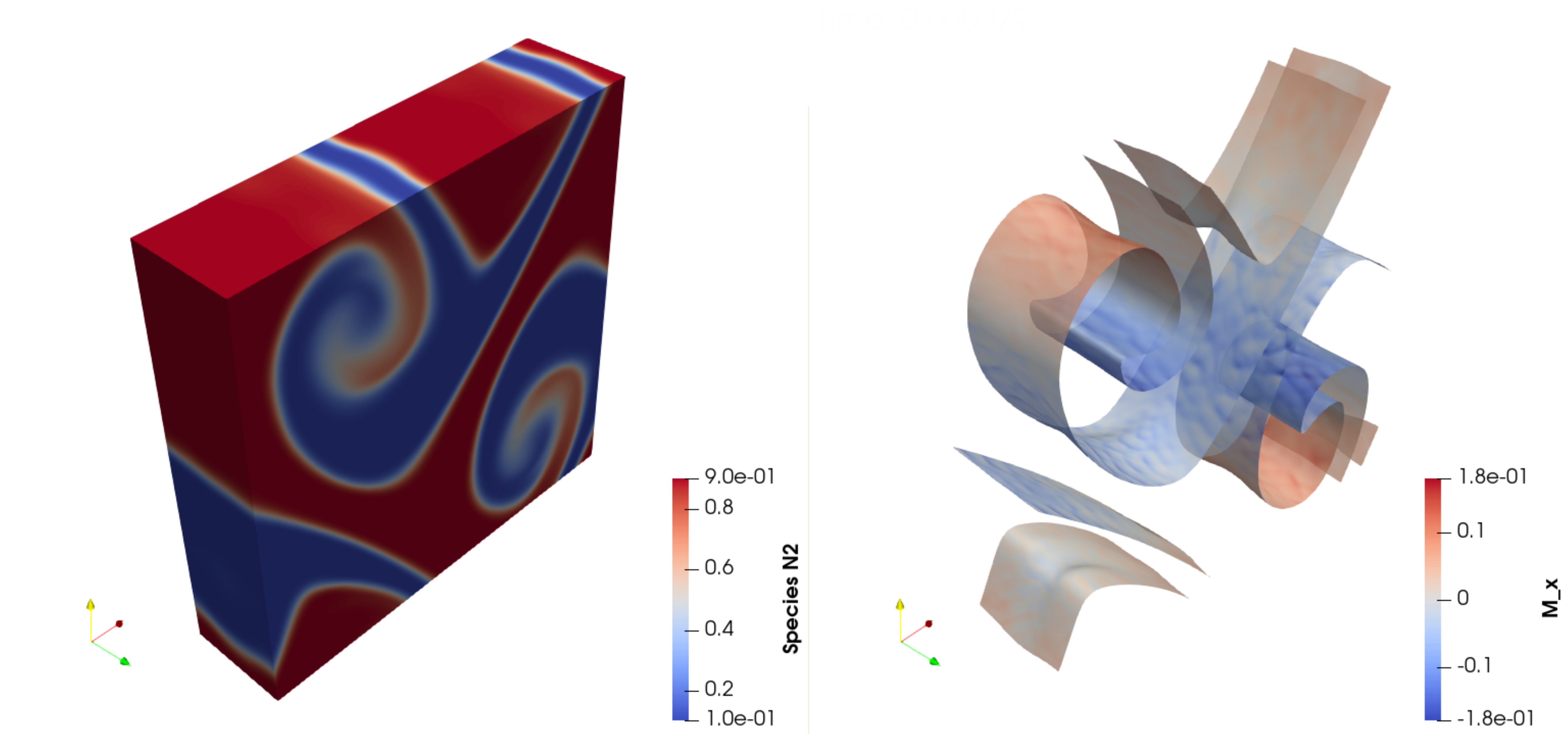}
 \caption{Contour of mole fraction of nitrogen $N_2$ (left) and iso-surface of the equilibrium concentration of nitrogen $X_{N_2}=0.5$ colored with the Mach number in the $x$-direction  (right) at time $t_e=3.3119$.}
 \label{fig:frame25_N2X}
\end{figure}
\begin{figure}
 \centering
 \includegraphics[keepaspectratio=true,width=0.9\textwidth]{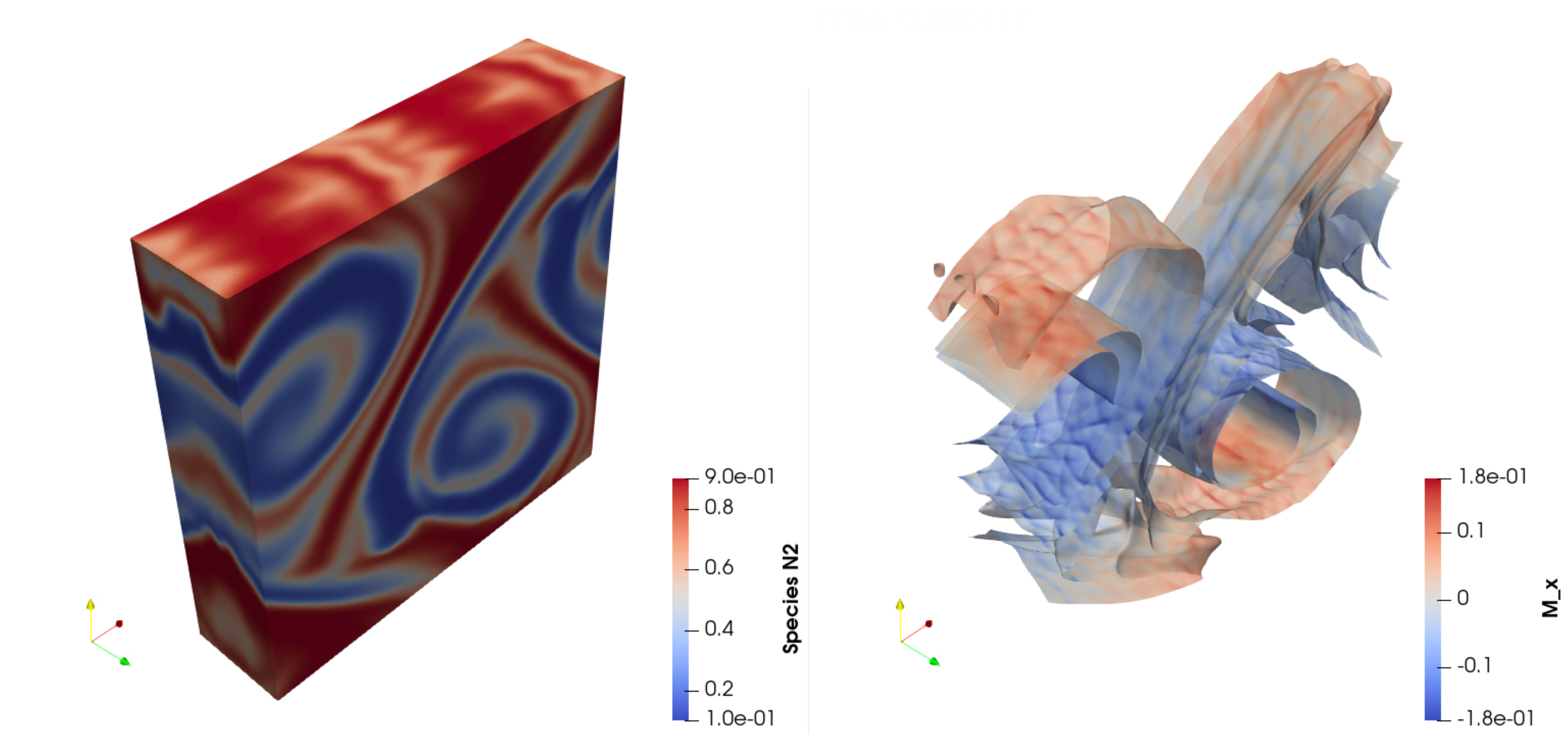}
 \caption{Contour of mole fraction of nitrogen $N_2$ (left) and iso-surface of the equilibrium concentration of nitrogen $X_{N_2}=0.5$ colored with the Mach number in the $x$-direction  (right) at time $t_e=4.901612$.}
 \label{fig:frame37_N2X}
\end{figure}
\begin{figure}
 \centering
 \includegraphics[keepaspectratio=true,width=0.9\textwidth]{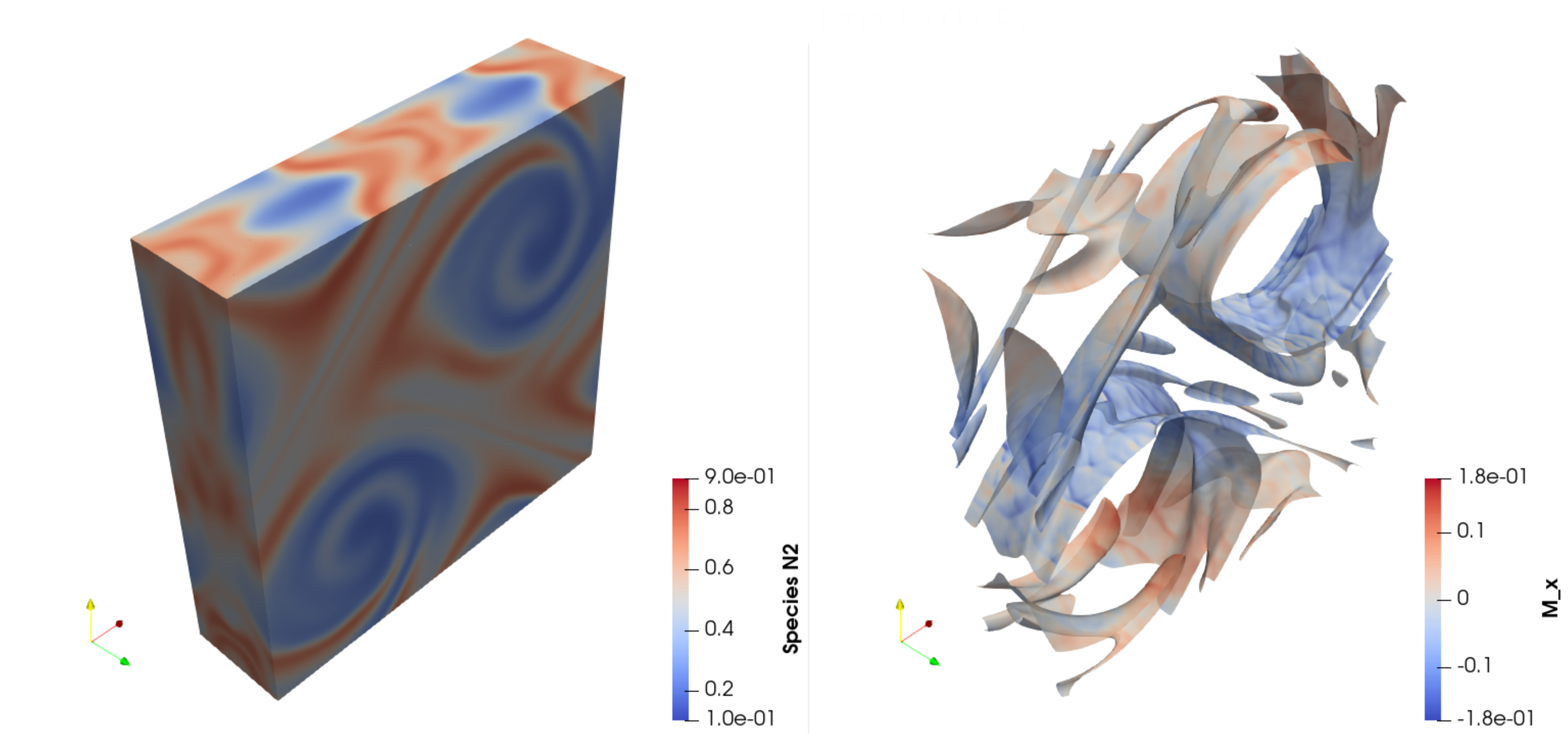}
 \caption{Contour of mole fraction of nitrogen $N_2$ (left) and iso-surface of the equilibrium concentration of nitrogen $X_{N_2}=0.5$ colored with the Mach number in the $x$-direction  (right) at time $t_e=8.345988$.}
 \label{fig:frame63_N2X}
\end{figure}

We present contours and iso-surfaces of the equilibrium mole fraction of nitrogen $X_{N_2}=0.5$ at different times  in Figs.\ \ref{fig:frame17_N2X} to \ref{fig:frame63_N2X}. 
The iso-surfaces are colored by the $x$-component of the convective Mach number $M_c$ to provide a visual indication of the direction of motion. 

\begin{figure}
 \centering
 \includegraphics[keepaspectratio=true,width=0.8\textwidth]{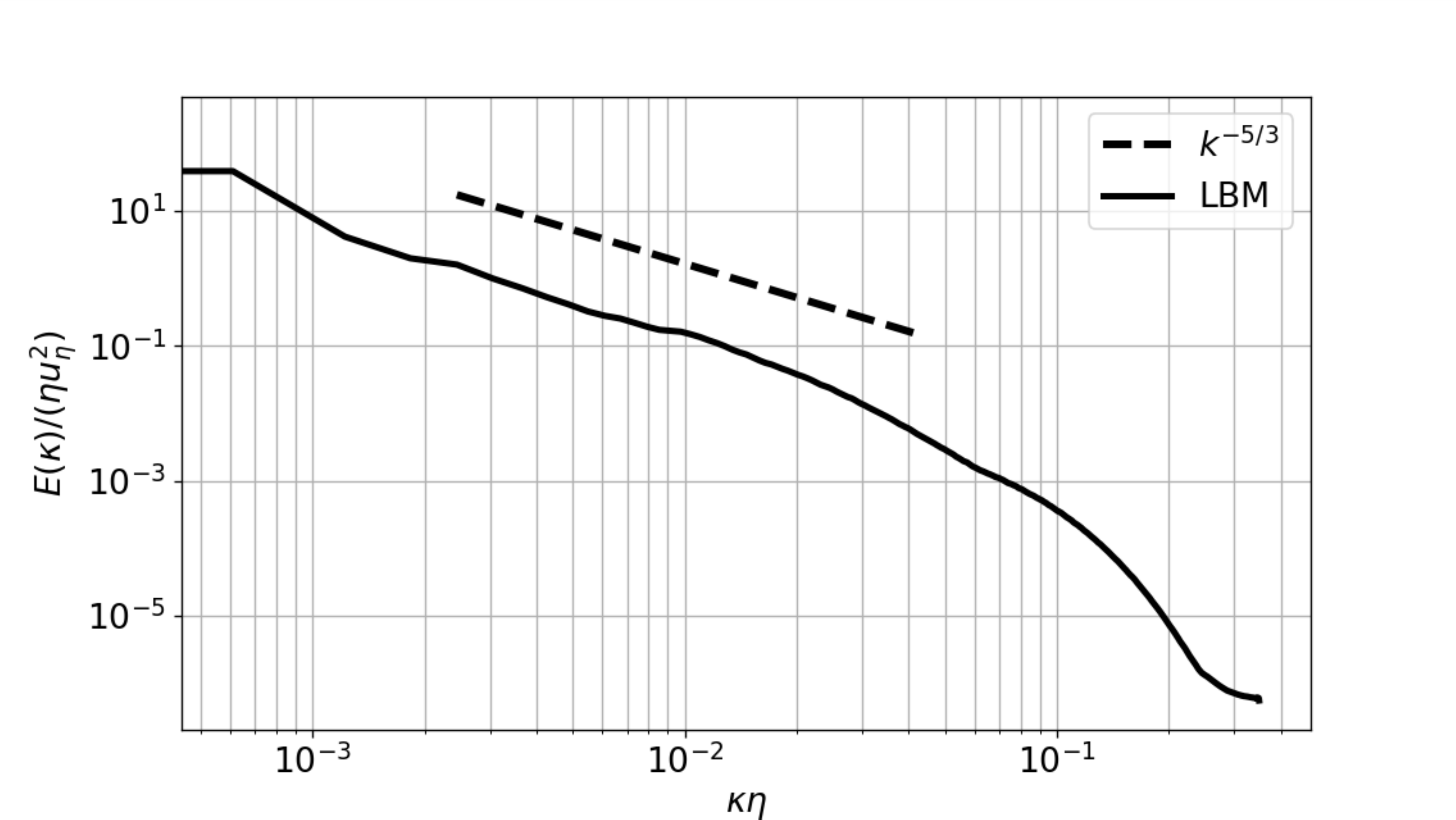}
 \caption{Turbulent kinetic energy spectrum at $t_e=7.9486$ along with the theoretical Kolmogorov scaling. Here $\eta$ is the Kolmogorov length scale and $u_{\eta}$ is the Kolmogorov velocity.}
 \label{fig:KHISpectrumKE}
\end{figure}

Soon after the initial condition, the normal perturbation breaks the symmetry of the flow and the shear layer begins to curl up into a vortex without a significant span-wise deformation. This is evident in Fig.\ \ref{fig:frame17_N2X}, where the three-dimensionality of the flow is visible only in the nonuniform span-wise velocity of the iso-surface. As the simulation proceeds, the flow in Fig.\ \ref{fig:frame25_N2X} develops anti-symmetric vortices which are also visibly deformed in the span-wise direction. This process continues and the vortices stretch and deform over time, which is visible in both the contours of the mole fraction of $N_2$ as well as the iso-surfaces in Fig.\ \ref{fig:frame37_N2X}. The flow eventually becomes more chaotic, forming smaller-scale structures in Fig.\ \ref{fig:frame63_N2X}. It should not be forgotten that the components are also undergoing diffusion during this mixing, as evident from the smearing of the contour values over time.   
%
%
While the above observations are inline with what is typically observed in
the literature, it is also important to verify the energy distribution 
across the scales of the flow. To that end, we measure the turbulent kinetic 
energy spectrum, which shows the expected $-5/3$ Kolmogorov scaling 
in the inertial subrange in Fig.\ \ref{fig:KHISpectrumKE}. 
This additionally validates our model and outlines a path towards complex 
multicomponent flow simulations.
%
%
\section{Conclusion}
\label{sec:conclusion}
Let us consider a ``good''  lattice Boltzmann setting for a single-component gas. From the past experience, this would imply a two-population LBM because the second population would be ultimately needed for an adequate description of the energy, leaving aside a special case of monatomic gas. 
It is then possible to envision, following the rule of the conventional kinetic theory \citep{chapmanCowling}, a valid lattice Boltzmann model for a mixture of such gases, with the total number of kinetic equations equal to $2\times M$ for the $M$-component mixture since each component needs to be represented by its individual two-population LBM.

On the contrary, in this paper we proposed a lattice Boltzmann framework for multicomponent mixtures of ideal gases with a more realistic number of coupled lattice Boltzmann equations. We addressed two equally important aspects. First, we proposed a new LBM system for the Stefan--Maxwell diffusion and barodiffusion comprising $M$ lattice Boltzmann equations. Second, we proposed a reduced, mean-field description of the mixture momentum and energy using the two-population setting. The resulting framework consists of $M+2$ lattice Boltzmann equations rather than $2\times M$ as it would be if the detailed and not the mean-field approach to the energy of the mixture would have been pursuit.

Special attention was devoted to the consistent thermodynamic coupling of the above two sub-systems in such a manner that the hydrodynamic limit is not compromised. The proposed framework was realized on the standard three-dimensional lattice using an extension of the compressible model of \citet{hosseinCompressible}. Specific to the multicomponent problem, the interdiffusion energy flux was added in a natural way to the heat flux to recover the correct energy equation while a counter-flux was introduced to remove the spurious contribution to the Fourier law inevitably arising with the mean-field approach to the energy description. 
While we focused on the multicomponent case, the proposed realization is also an extension of the augmented, compressible LB model 
proposed by \citet{hosseinCompressible} to a general form in three dimensions, which can of course also be used for single-component flows.

The simulation of the diffusion in a ternary mixture demonstrated that the proposed LBM correctly accounts for binary interaction between species. The coupling of diffusion to hydrodynamics 
was assessed by computing diffusion in opposed jets and the basic compressibility features were demonstrated through the speed of sound simulation at various compositions. Finally, the simulation of the three-dimensional shear layer instability in a binary mixture with a high composition contrast indicates that the proposed method can be useful for direct numerical simulations of complex flows. 

All of the above gives us grounds to believe that the proposed multicomponent framework fills the gap in the development of the lattice Boltzmann method and is a first step towards reactive flow applications which will be the focus of our future studies.

\noindent {\bf Acknowledgement}. 
This work was supported by European Research Council (ERC) Advanced Grant No. 834763-PonD. 
Computational resources at the Swiss National  Super  Computing  Center  CSCS  were  provided  under grant No. s897. 
Authors thank S. Springman and A. Togni for enabling research of N. S. at ETHZ.

\noindent {\bf Declaration of interests}.
The authors report no conflict of interest.

\appendix

\section{Hydrodynamic limit of the mean-field LBM}
\label{sec:ceNavierStokes}
%
We expand the lattice Boltzmann equations (\ref{eqn:f}) and (\ref{eqn:g}) in Taylor series to second order, using space component notation and summation convention,
\begin{align}
\left[\delta t (\partial_t + c_{i\mu} \partial_\mu ) + \frac{\delta t^2}{2} (\partial_t + c_{i\mu} \partial_\mu )^2\right] f_i &= \omega (f_i^{\rm eq} -f_i),
\label{eqn:ftaylor} 
\\
\left[\delta t (\partial_t + c_{i\mu} \partial_\mu ) + \frac{\delta t^2}{2} (\partial_t + c_{i\mu} \partial_\mu )^2\right] g_i &= \omega_1 (g_i^{\rm eq} -g_i) + (\omega - \omega_1) (g_i^* -g_i).
\label{eqn:gtaylor}
\end{align}
With a time scale $\bar t$ and a velocity scale $\bar c$, the non-dimensional parameters are introduced  as follows,
\begin{equation}
t'=\frac{t}{\bar t}, \; c_{\alpha}'=\frac{c_{\alpha}}{\bar c}, \; x_{\alpha}'=\frac{x_{\alpha}}{\bar c \bar t}.
\label{eqn:nd} 
\end{equation}
Substituting the relations (\ref{eqn:nd}) into  (\ref{eqn:ftaylor}) and (\ref{eqn:gtaylor}), the kinetic equations in the non-dimensional form become,
\begin{align}
\left[\delta t' (\partial_{t'} + c'_{i\mu} \partial_{\mu'} ) + \frac{\delta t'^2}{2} (\partial_{t'} + c'_{i\mu} \partial_{\mu'} )^2\right] f_i &= \omega (f_i^{\rm eq} -f_i),
\label{eqn:ftaylorNonDimensional} 
\\
\left[\delta t' (\partial_{t'} + c'_{i\mu} \partial_{\mu'} ) + \frac{\delta t'^2}{2} (\partial_{t'} + c'_{i\mu} \partial_{\mu'} )^2\right] g_i &= \omega_1 (g_i^{\rm eq} -g_i) + (\omega - \omega_1) (g_i^* -g_i).
\label{eqn:gtaylorNonDimensional}
\end{align}
Let us define a smallness parameter $\epsilon$ as,
\begin{equation}
\epsilon=\delta t'=\frac{\delta t}{\bar t}.
\label{eqn:defineEpsillon} 
\end{equation}
Using the definition of $\epsilon$ and dropping the primes for ease of writing, we obtain,
\begin{align}
\left[\epsilon (\partial_t + c_{i\mu} \partial_\mu ) + \frac{\epsilon^2}{2} (\partial_t + c_{i\mu} \partial_\mu )^2\right] f_i &= \omega (f_i^{\rm eq} -f_i),
\label{eqn:feps} 
\\
\left[\epsilon (\partial_t + c_{i\mu} \partial_\mu ) + \frac{\epsilon^2}{2}(\partial_t + c_{i\mu} \partial_\mu )^2\right] g_i &= \omega_1 (g_i^{\rm eq} -g_i) + (\omega - \omega_1) (g_i^* -g_i).
\label{eqn:geps}
\end{align}
Writing a power series expansion in $\epsilon$ as,
\begin{align}
\partial_t &= \partial_t^{(1)} + \epsilon \partial_t^{(2)},
\label{eqn:epst} 
\\
f_{i} &= f_{i}^{(0)} + \epsilon f_{i}^{(1)} + \epsilon^2 f_{i}^{(2)},
\label{eqn:epsf} 
\\
g_{i} &= g_{i}^{(0)} + \epsilon g_{i}^{(1)} + \epsilon^2 g_{i}^{(2)},
\label{eqn:epsg} 
\\
g_{i}^* &= g_{i}^{*(0)} + \epsilon g_{i}^{*(1)} + \epsilon^2 g_{i}^{*(2)},
\label{eqn:epsgstar} 
\end{align}
we substitute the equations (\ref{eqn:epst}--\ref{eqn:epsgstar}) into (\ref{eqn:feps}) and (\ref{eqn:geps}), and proceed with collecting terms of same order. This Chapman--Enskog analysis is standard \citep{chapmanCowling}; for the specific case of the two-population LBM see, e.\ g., \citep{karlinConsistent}. At order $\epsilon^0$, we get,
\begin{align}
f_i^{(0)} &= f_i^{\rm eq},
\label{eqn:f0feq} 
\\
g_i^{(0)} &= g_i^{*(0)} = g_i^{\rm eq}.
\label{eqn:g0geq} 
\end{align}
At order $\epsilon^1$, upon summation over the discrete velocities, we find,
\begin{align}
&\partial_t^{(1)} \rho + \partial_{\alpha} j_\alpha^{\rm eq} = 0,
\label{eqn:dt1rhoeps1}
\\
&\partial_t^{(1)} j_\alpha^{\rm eq} + \partial_\beta P_{\alpha \beta}^{\rm eq} = 0,
\label{eqn:dt1ueps1}
\\
&\partial_t^{(1)} (\rho E) + \partial_{\alpha} q_\alpha^{\rm eq} = 0.
\label{eqn:dt1Teps1}
\end{align}
Here, $\rho$ is the density of the fluid given by the zeroth moment of the $f$-populations in equation (\ref{eqn:f0mom}), $j_\alpha^{\rm eq}$ is the equilibrium momentum of the fluid as defined by equation (\ref{eqn:f1mom}), $P_{\alpha \beta}^{\rm eq}$ is the equilibrium pressure tensor and $q_\alpha^{\rm eq}$ is the equilibrium heat flux as defined by equations (\ref{eqn:feq2mom}) and (\ref{eqn:geq1mom}), respectively, and  $\rho E$ is the total energy of the fluid calculated as the zeroth moment of $g$-populations using equation (\ref{eqn:g0mom}).
Finally, at order $\epsilon^2$ we arrive at,
\begin{align}
&\partial_t^{(2)} \rho = 0,
\label{eqn:dt2rhoeps2}\\
%
&\partial_t^{(2)} j_\alpha^{\rm eq} + \left( \frac{1}{2} - \frac{1}{\omega}  \right) \partial_\beta (\partial_t^{(1)} P_{\alpha \beta}^{\rm eq} + \partial_\gamma Q_{\alpha \beta \gamma}^{\rm eq})=0,
\label{eqn:dt2ueps2}\\
%
&\partial_t^{(2)} (\rho E )+ \partial_\alpha \left[ \left( \frac{1}{2} - \frac{1}{\omega}\right) (\partial_t^{(1)} q_\alpha^{\rm eq} + \partial_\beta R_{\alpha \beta}^{\rm eq}) + \left( 1 -\frac{\omega_1}{\omega}\right) q_\alpha^{*(1)} \right] =0.
\label{eqn:dt2Teps2}
\end{align}
Here, $Q_{\alpha \beta \gamma}^{\rm eq}$ and $R_{\alpha \beta}^{\rm eq}$ are the third-order moment of $f_i^{\rm eq}$ and the second-order moment of $g_i^{\rm eq}$, respectively; Their expressions are given by (\ref{eqn:feq3mom}) and  (\ref{eqn:geq2mom}), respectively.
Combining terms at both orders, we recover the following macroscopic equations,
\begin{align}
&\partial_t \rho + \partial_{\alpha} j_\alpha^{\rm eq}=0,
\label{eqn:dtRhoAppendix}\\
&\partial_t j_\alpha^{\rm eq} + \partial_\beta P_{\alpha \beta}^{\rm eq} +
 \epsilon \left( \frac{1}{2} - \frac{1}{\omega}  \right) \partial_\beta (\partial_t^{(1)} P_{\alpha \beta}^{\rm eq} + \partial_\gamma Q_{\alpha \beta \gamma}^{\rm eq})=0,
\label{eqn:dtRhoUAppendix}\\
&\partial_t (\rho E) + \partial_{\alpha} q_\alpha^{\rm eq} + 
\epsilon  \partial_\alpha \left[ \left( \frac{1}{2} - \frac{1}{\omega}\right) (\partial_t^{(1)} q_\alpha^{\rm eq} + \partial_\beta R_{\alpha \beta}^{\rm eq}) + \left( 1 -\frac{\omega_1}{\omega}\right) q_\alpha^{*(1)} \right]  =0,
\label{eqn:dtRhoEAppendix}
\end{align}
where, 
\begin{align}
&\partial_t^{(1)} P_{\alpha \beta}^{\rm eq} + \partial_\gamma Q_{\alpha \beta \gamma}^{\rm eq} = -\frac{PR}{C_v} \partial_\gamma u_\gamma \delta_{\alpha \beta}+ P (\partial_\alpha u_\beta + \partial_\beta u_\alpha), 
\label{eqn:dt1P} \\
&\partial_t^{(1)} q_\alpha^{\rm eq} + \partial_\beta R_{\alpha \beta}^{\rm eq} = P {\left(1-\frac{C_p}{C_v}\right)} u_\alpha \partial_\beta u_\beta + P u_\beta (\partial_\alpha u_\beta + \partial_\beta u_\alpha)
{+ P \sum_{a=1}^M H_a\partial_{\alpha}Y_a+PC_p\partial_{\alpha}T,}
\label{eqn:dt1q} \\
&q_\alpha^{*(1)}= \left( \frac{1}{\omega_1} \right) (\partial_t^{(1)} q_\alpha^{\rm eq} + \partial_\beta R_{\alpha \beta}^{\rm eq}) + \frac{1}{\epsilon} \left( \frac{\omega}{\omega_1} \right) (- u_\beta (P_{\alpha \beta}-P_{\alpha \beta}^{\rm eq}) + q_\alpha^{\rm diff} + q_\alpha^{\rm corr}) ,
\label{eqn:qStar1} \\
 & P_{\alpha \beta}-P_{\alpha \beta}^{\rm eq} = \epsilon \left(-\frac{1}{\omega}\right) (\partial_t^{(1)} P_{\alpha \beta}^{\rm eq} + \partial_\gamma Q_{\alpha \beta \gamma}^{\rm eq}).
\label{eqn:Pneq}
\end{align}
We now substitute for the moments from the expressions (\ref{eqn:dt1P}) to (\ref{eqn:Pneq}) in equations (\ref{eqn:dtRhoAppendix}) to (\ref{eqn:dtRhoEAppendix}) and for the equilibrium moments from equations (\ref{eqn:f0mom}--\ref{eqn:gstareq2mom}) to get the resulting macroscopic equations.
Equation (\ref{eqn:dtRhoAppendix}) recovers the continuity equation,
\begin{align}
&\partial_t \rho + \partial_\alpha (\rho u_\alpha) = 0.
\label{eqn:continuityMixAppendix} 
\end{align}
Equation (\ref{eqn:dtRhoUAppendix}) recovers the mixture momentum equation,
\begin{align}
&\partial_t (\rho u_\alpha) + \partial_\beta (\rho u_\alpha u_\beta) + \partial_\beta \pi_{\alpha \beta}= 0,
\label{eqn:momentumMixAppendix} 
\end{align}
with the constitutive relation for the stress tensor,
\begin{align}
\pi_{\alpha \beta} = P \delta_{\alpha \beta} - \mu 
\left(\partial_\alpha u_\beta + \partial_\beta u_\alpha- \frac{2}{D} (\partial_\mu u_\mu) \delta_{\alpha \beta}\right) - \varsigma (\partial_\mu u_\mu) \delta_{\alpha \beta}.
\label{eqn:piMix}
\end{align}
The dynamic viscosity $\mu$ and the bulk viscosity $\varsigma$ are related to the relaxation coefficient $\omega$ by equations (\ref{eq:mu}) and (\ref{eqn:varsigma}), respectively.
Finally, equation (\ref{eqn:dtRhoEAppendix}) recovers the mixture energy equation,
%
%
%
\begin{align}
\label{eqn:energyMixIntermediateAppendix}
\partial_t (\rho E) + \partial_\alpha(\rho Eu_\alpha)+ \partial_\alpha ( \pi_{\alpha \beta}u_\beta) +\partial_{\alpha}q_{\alpha}=0,
\end{align}
where the heat flux $\bm{q}$ has the following form,
\begin{align}
\label{eqn:energyFluxIntermediateAppendix}
q_{\alpha}=-\lambda\partial_\alpha T - {{ \epsilon P \left( \frac{1
}{\omega_1}- \frac{1}{2} \right)\sum_{a=1}^M H_a \partial_\alpha Y_a } }+ {\left( \frac{\omega}{\omega_1}-1 \right) q_\alpha^{\rm corr}} + \left( \frac{\omega}{\omega_1}-1 \right) q_\alpha^{\rm diff},
\end{align}
with  the thermal conductivity $\lambda$ defined by equation (\ref{eq:lambda}). 
We now chose $q_\alpha^{\rm corr}$ to cancel the spurious second term containing the gradient of {$Y_a$},
\begin{align}
q_\alpha^{\rm corr} = \frac{1}{2} \left( \frac{\omega_1-2}{\omega_1-\omega} \right) \epsilon  P {\sum_{a=1}^M H_a \partial_\alpha Y_a .}
\label{eqn:qcorr}
\end{align}
This is equivalent to Eq.\ (\ref{eq:corrFourier}). Finally, the inter-diffusion energy flux is introduced by choosing the last term $\bm{q}^{\rm diff}$ in (\ref{eqn:energyFluxIntermediateAppendix}) as,
\begin{align}
q_\alpha^{\rm diff} = \left( \frac{\omega_1}{\omega-\omega_1} \right)  \rho \sum_{a=1}^{M} H_a Y_a V_{a\alpha},
\label{eqn:qdiff} 
\end{align}
which is equivalent to Eq.\ (\ref{eq:interdiffusion}).
Substituting (\ref{eqn:qcorr}) and (\ref{eqn:qdiff}) into (\ref{eqn:energyFluxIntermediateAppendix}), we get the heat flux ${\bm{q}}$ in the energy equation (\ref{eqn:energyMixIntermediateAppendix}) as a combination of the Fourier law and the inter-diffusion energy flux due to diffusion of the species \citep{williams,bird2006transport},
%
\begin{equation}
{q_\alpha} = - \lambda \partial_\alpha T +  \rho \sum_{a=1}^{M} H_a Y_a V_{a\alpha}.   
\label{eqn:qMixCE} 
\end{equation}

\bibliographystyle{jfm}
\bibliography{diffusionText.bib}

\end{document}